\documentclass[twocolumn]{article}
\usepackage[hmargin=0.65in,vmargin=0.7in,columnsep=0.2in]{geometry}
\usepackage[utf8]{inputenc}
\usepackage{amsmath}
\usepackage{authblk}
\usepackage{upgreek}
\usepackage{siunitx}
\usepackage{graphicx}
\usepackage{braket}
\usepackage[version=4]{mhchem}
\usepackage{bbm}
\usepackage{color}
\usepackage[sort,square,compress,comma,numbers]{natbib}
\usepackage{hyperref}
\usepackage[normalem]{ulem}

\newcommand{\pbrac}[1]{\left(#1\right)}

\DeclareMathOperator{\id}{\mathbbm{1}}

\definecolor{JM}{RGB}{4,116,149}

\title{Quantum circuits with many photons \\ on a programmable nanophotonic chip}
\author[1]{J.M. Arrazola}
\author[1]{V. Bergholm}
\author[1]{K. Br{\'a}dler}
\author[1]{T.R. Bromley}
\author[1]{M.J. Collins}
\author[1]{I. Dhand}
\author[1]{A. Fumagalli}
\author[2]{T. Gerrits}
\author[1]{A. Goussev}
\author[1]{L.G. Helt}
\author[1]{J. Hundal}
\author[1]{T. Isacsson}
\author[1]{R.B. Israel}
\author[1]{J. Izaac}
\author[1]{S. Jahangiri}
\author[1]{R. Janik}
\author[1]{N. Killoran}
\author[1]{S.P. Kumar}
\author[1]{J. Lavoie}
\author[2]{A.E. Lita}
\author[1]{D.H. Mahler}
\author[1]{M. Menotti}
\author[1]{B. Morrison}
\author[2]{S.W. Nam}
\author[1]{L. Neuhaus}
\author[1]{H.Y. Qi}
\author[1]{N. Quesada}
\author[1]{A. Repingon}
\author[1]{K.K. Sabapathy}
\author[1]{M. Schuld}
\author[1]{D. Su}
\author[1]{J. Swinarton}
\author[1]{A. Sz{\'a}va}
\author[1]{K. Tan}
\author[1]{P. Tan}
\author[1]{V.D. Vaidya}
\author[1]{Z. Vernon}
\author[1]{Z. Zabaneh}
\author[1]{Y. Zhang}

\affil[1]{Xanadu, Toronto, ON, M5G 2C8, Canada}
\affil[2]{National Institute of Standards and Technology, Boulder, CO, USA}

\date{}

\begin{document}
\maketitle 
\section*{Abstract} Growing interest in quantum computing for practical applications has led to a surge in the availability of programmable machines for executing quantum algorithms. Present day photonic quantum computers have been limited either to non-deterministic operation, low photon numbers and rates, or fixed random gate sequences. Here we introduce a full-stack hardware-software system for executing many-photon quantum circuits using integrated nanophotonics: a programmable chip, operating at room temperature and interfaced with a fully automated control system. It enables remote users to execute quantum algorithms requiring up to eight modes of strongly squeezed vacuum initialized as two-mode squeezed states in single temporal modes, a fully general and programmable four-mode interferometer, and genuine photon number-resolving readout on all outputs. Multi-photon detection events with photon numbers and rates exceeding any previous quantum optical demonstration on a programmable device are made possible by strong squeezing and high sampling rates. We verify the non-classicality of the device output, and use the platform to carry out proof-of-principle demonstrations of three quantum algorithms: Gaussian boson sampling, molecular vibronic spectra, and graph similarity.

\section*{Introduction}
The last decade has seen remarkable progress in quantum computation and simulation. Breakthroughs across a diverse range of physical platforms have enabled the construction of programmable machines that can deliver the automation, stability, and repeatability demanded by increasingly sophisticated quantum algorithms. Rigorous benchmarks have been carried out successfully on an 11-qubit trapped ion system~\cite{wright2019benchmarking, kielpinski2002architecture}, and a 53-qubit superconducting system has been used to generate random samples from a probability distribution at a rate exceeding what is reasonably achievable using classical hardware~\cite{arute2019quantum,clarke2008superconducting}. Similar machines can now be remotely accessed and loaded with algorithms written in high-level programming languages by users having no intimate knowledge of the low-level quantum hardware details of the apparatus. These capabilities have rapidly accelerated research targeting application development for near-term quantum computers~\cite{wootton2018repetition,dumitrescu2018cloud,anschuetz2019variational}.

Such hardware has primarily been designed to access problems in the qubit model, where computation is carried out by initializing a quantum state in a space spanned by a product of binary-valued basis states, and performing a sequence of quantum gates selected from a typically discrete set of operations~\cite{nielsen2002quantum}. Present-day examples of these machines, however, are limited to dozens of noisy qubits, restricting their applicability to quantum algorithms that are compatible with this scale \cite{preskill2018quantum}. Other algorithms are more efficiently expressed in a model in which each independent quantum system is described by a state in an infinite-dimensional Hilbert space. Examples of such applications include those implementing bosonic error correction codes~\cite{gottesman2001encoding,fluhmann2019encoding}, a wide class of Gaussian boson sampling applications~\cite{huh2015boson,arrazola2018using,bradler2018graph,bradler2017gaussian,schuld2020measuring,banchi2019molecular,bromley2020applications}, and other bespoke algorithms that exploit the mathematical structure of infinite-dimensional Hilbert spaces~\cite{killoran2019continuous, arrazola2019quantum}. 

A promising platform for the large-scale implementation of such bosonic quantum algorithms is offered by photonic hardware. A number of groundbreaking demonstrations of photonic quantum information processing have recently been carried out. Two-dimensional cluster states with tens of thousands of entangled nodes have been deterministically generated using free-space and fiber-optical components~\cite{larsen2019deterministic,asavanant2019generation}, and photonic experiments have been constructed to sample from the photon number distribution of multi-mode Gaussian states~\cite{paesani2019generation,zhong2019experimental}. Combined with rapid advancements in photonic chip fabrication~\cite{wang2019integrated}, such demonstrations coincide with new optimism towards photonics as a platform for advancing the frontier of quantum computation~\cite{rudolph2017optimistic}. 

Despite these advances, much work remains in developing photonic systems for practical use in quantum computation. Large photonic cluster state demonstrations~\cite{larsen2019deterministic,asavanant2019generation} were limited to all-Gaussian states, gates, and measurements, rendering them efficiently simulable at any scale by classical computers. Single-photon-based experiments on integrated platforms~\cite{qiang2018large} suffer from non-deterministic state preparation and gate implementation, severely hindering their scalability. This deficit can be evaded in photonic experiments by using deterministically-prepared squeezed states and linear optics, with non-Gaussian operations provided by photon-counting detectors. In such experiments, and in the machine we present, the squeezed state inputs play the role of qubits as the basic independently accessible quantum systems. But demonstrations of such squeezing-based photonic machines~\cite{paesani2019generation,zhong2019experimental} lacked programmability, with each accessing only a fixed, randomized quantum state. Furthermore, these demonstrations were limited to small numbers of detected photons. 

To date, no photonic machine has been demonstrated that is simultaneously (i) dynamically programmable, (ii) readily scalable to hundreds of modes and photons, and (iii) able to access a class of quantum circuits that could not, when the system size is scaled, be efficiently simulated by classical hardware. Here we report results from a new device based on a programmable nanophotonic chip which includes all of these capabilities in a single scalable and unified machine. We describe the functional performance of the components designed for initial state preparation, gate sequence implementation, and readout, and we verify the non-classicality of the device output. We then use the machine to carry out proof-of-principle demonstrations of the execution of three types of quantum algorithms: Gaussian boson sampling \cite{hamilton2017gaussian}, molecular vibronic spectra \cite{huh2015boson}, and graph similarity \cite{schuld2020measuring}. The Gaussian boson sampling experiment has the largest number of detected photons to date, and the graph similarity demonstration is the first of its kind. While our device, at its current scale, can be readily simulated by a classical computer, the architecture and platform developed can potentially enable future generations of such machines to exit this regime and perform tasks that are not practically simulable by classical systems.

\section*{Hardware}
As an overview of the hardware platform developed, we summarize the layout of the chip and other core subsystems, and demonstrate the performance of the key quantum components by measuring three vital metrics: noise reduction factors, second-order correlation statistics, and multi-photon interference between distinct sources. We then verify the non-classicality of the device output.

\subsection*{Chip and control system}
\begin{figure}
    \centering
    \includegraphics[width=\columnwidth]{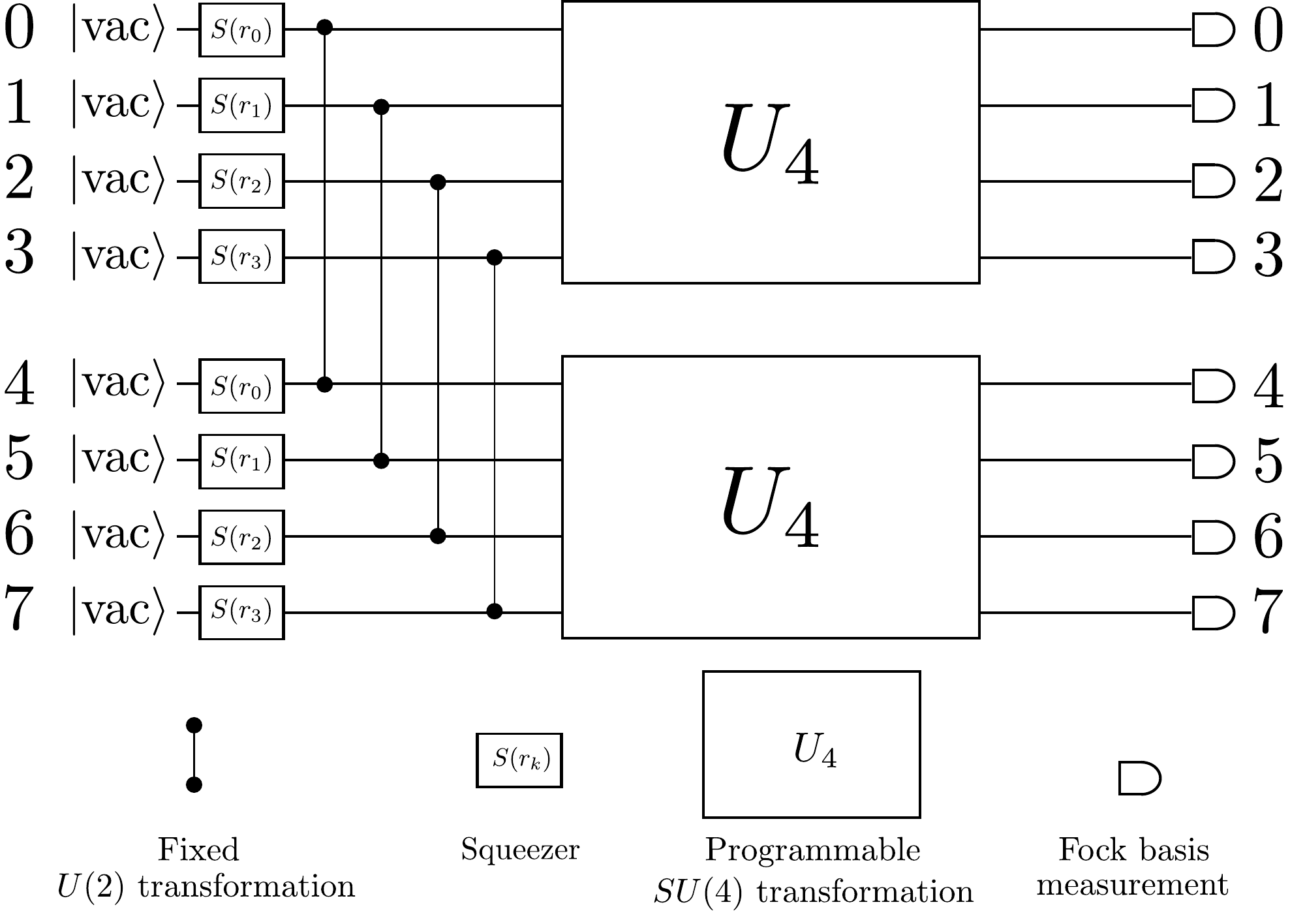}
    \caption{Equivalent quantum circuit diagram illustrating the functionality of the photonic hardware. Up to eight modes initialized as vacuum are squeezed with squeezing parameters $r_k$ and entangled (via the fixed two-mode $U(2)$ transformation equivalent to a 50/50 beam splitter with the relative input phase set to produce two-mode squeezing at the output) to form two-mode squeezed vacuum states. Programmable four-mode rotation gates ($SU(4)$ transformation, represented by the large boxes labelled $U_4$ in the figure) are applied to each four-mode subspace. All eight modes are individually read out by measurements in the Fock basis.}
    \label{fig:circuit_diagram}
\end{figure}

The essential component of our device is a $10$ mm $\times 4$ mm photonic chip. It generates squeezed light~\cite{lvovsky2015squeezed} in up to eight separate optical modes, with a fixed initialization into four independent two-mode squeezed vacuum states. The two-mode squeezing is generated between bichromatic mode pairs, with each such pair populating one of four spatially-separated waveguide modes. An interferometer, based on a network of beam splitters and phase shifters, implements a user-programmable gate sequence corresponding to an $SU(4)$ transformation (with $SU(n)$ the special unitary group of degree $n$) applied to the spatial modes. The user must specify twelve independent real parameters to program this transformation, with the remaining three free parameters of the $SU(4)$ transformation corresponding to irrelevant output phases. This transformation implements the gate sequence on both four-mode subspaces distinguished by their optical wavelength. The resultant eight-mode programmable Gaussian state synthesized by the chip is then measured in the Fock basis using eight independent photon number-resolving (PNR) detectors. An equivalent quantum circuit diagram for the machine is illustrated in Fig.~\ref{fig:circuit_diagram}.

\begin{figure}[t!]
    \centering
    \includegraphics[width=\columnwidth]{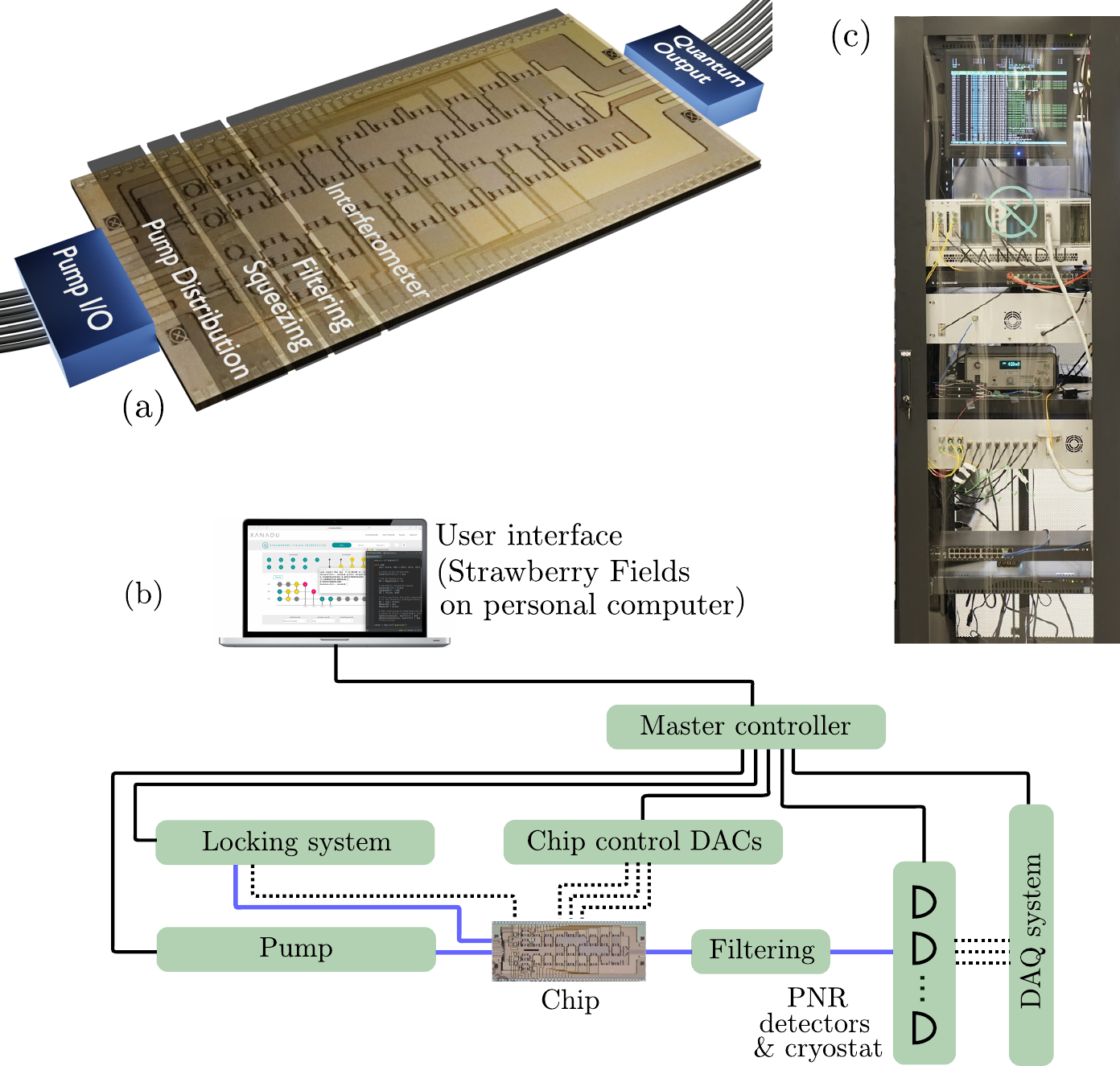}
    \caption{(a) Rendering of the chip (based on micrograph of true device) showing fiber optical inputs and outputs, and on-chip modules for coherent pump power distribution, squeezing, pump filtering, and programmable linear optical transformations. (b) Schematic of full apparatus and control system. Solid (dashed) black lines indicate digital (analog) electronic signals; blue lines indicate optical signals. DAC: digital to analog converter; DAQ: data acquisition; PNR: photon number resolving. (c) Photograph of entire system (except for PNR detector hardware), which has been fitted into a standard server rack.}
    \label{fig:apparatus}
\end{figure}

The full apparatus is illustrated in Fig.~\ref{fig:apparatus}. The chip itself (Fig.~\ref{fig:apparatus}(a)) is based on silicon nitride waveguides and thermo-optic phase shifters. It was fabricated using a photolithographic process on a dedicated wafer run through a commercial service offered by Ligentec SA. The die contains modules for coherent distribution of pump light, generation of squeezed states, filters to separate pump light from generated quantum signals, and programmable linear-optical transformations. Four squeezers based on microring resonators~\cite{vaidya2019broadband} are integrated, each of which generates a bichromatic two-mode squeezed state in a nearly-single temporal mode when pumped with a pulsed laser; equivalently stated, each squeezer generates an entangled mode pair in its respective waveguide output. The modes in these pairs are distinguished by wavelength, and we refer to them as the ``signal" and ``idler", in keeping with standard terminology. Four asymmetric Mach-Zehnder interferometer (AMZI) filters separate the pump light from the generated squeezed states, directing the squeezed light into the programmable interferometer, and the pump light out of the chip. This remaining pump light is monitored off-chip to provide a stabilization signal for the squeezer resonators. The interferometer implements an arbitrary programmable four-mode linear optical transformation on both the signal and idler subspaces of the squeezed light. The use of two-mode squeezers doubles the total number of modes available for detection per spatial mode, at the cost of restricting the space of eight-mode Gaussian states accessible from the chip. The synthesized Gaussian state is then coupled out of the chip for photon counting. More detail is provided in the Methods section.

To operate the apparatus, a control system was developed to autonomously actuate all required control signals, monitor system status, and acquire data. An overview diagram of the full system is shown in Fig.~\ref{fig:apparatus}(b). A master controller (conventional server computer) running custom-developed control software coordinates the operation of the chip and all other hardware required, which includes: (i) a custom modulated pump laser source, (ii) an active locking system for the on-chip squeezer resonators, (iii) digital-to-analog converters for tuning the AMZI filters and programming the interferometer, (iv) a cryogenic system (adiabatic demagnetization refrigerator) hosting the photon counting detectors, (v) automated detector control electronics, and (vi) a real-time data acquisition system for detector readout. The system is accessed by a high-level application programming interface: a classical computer providing the quantum programs for the photonic chip, using the Strawberry Fields Python library~\cite{killoran2019strawberry}. This enables users with no knowledge of the hardware details to remotely run quantum algorithms on the device. Apart from the photon counting system, the entire machine is contained in a standard server rack (Fig.~\ref{fig:apparatus}(c)); the chip itself is optically and electronically packaged, forming a mechanically stable solid-state system. The full apparatus is alignment-free and indefinitely stable for continuous operation, except for the cryogenic detection system, which requires two hours of downtime every 24 hours for its automated cycling process to complete. 

In contrast with demonstrations of earlier photonic devices~\cite{paesani2019generation,zhong2019experimental}, our machine features non-classical light sources which are designed to generate squeezed light in single temporal modes with high average photon number (squeezing parameter $r\approx 1$, mean photon number  $\overline{n}=\sinh^2(r)\approx 1.4$ at the sources). In addition, detection is carried out using transition edge sensors, yielding true photon number resolution at the readout stage \cite{rosenberg2005noise}. This enables the execution of quantum algorithms that involve strong multi-photon contributions, a key requirement for implementing a wide range of squeezing-based photonic quantum applications. For example, large photon number contributions are essential for accessing higher energy transitions when using a photonic device for vibronic spectrum simulations~\cite{huh2015boson}. Achieving high photon numbers is also crucial for achieving a quantum advantage~\cite{qi2020regimes}. Our device readily achieves large photon number event rates exceeding all previous demonstrations of programmable photonic devices: with all squeezers activated, four-photon detection events occur at an average event rate of $10\,000$ events/s, ten-photon events at $270$ events/s, and nineteen-photon events at $0.3$ events/s.

\begin{figure}
    \centering
    \includegraphics[width=\columnwidth]{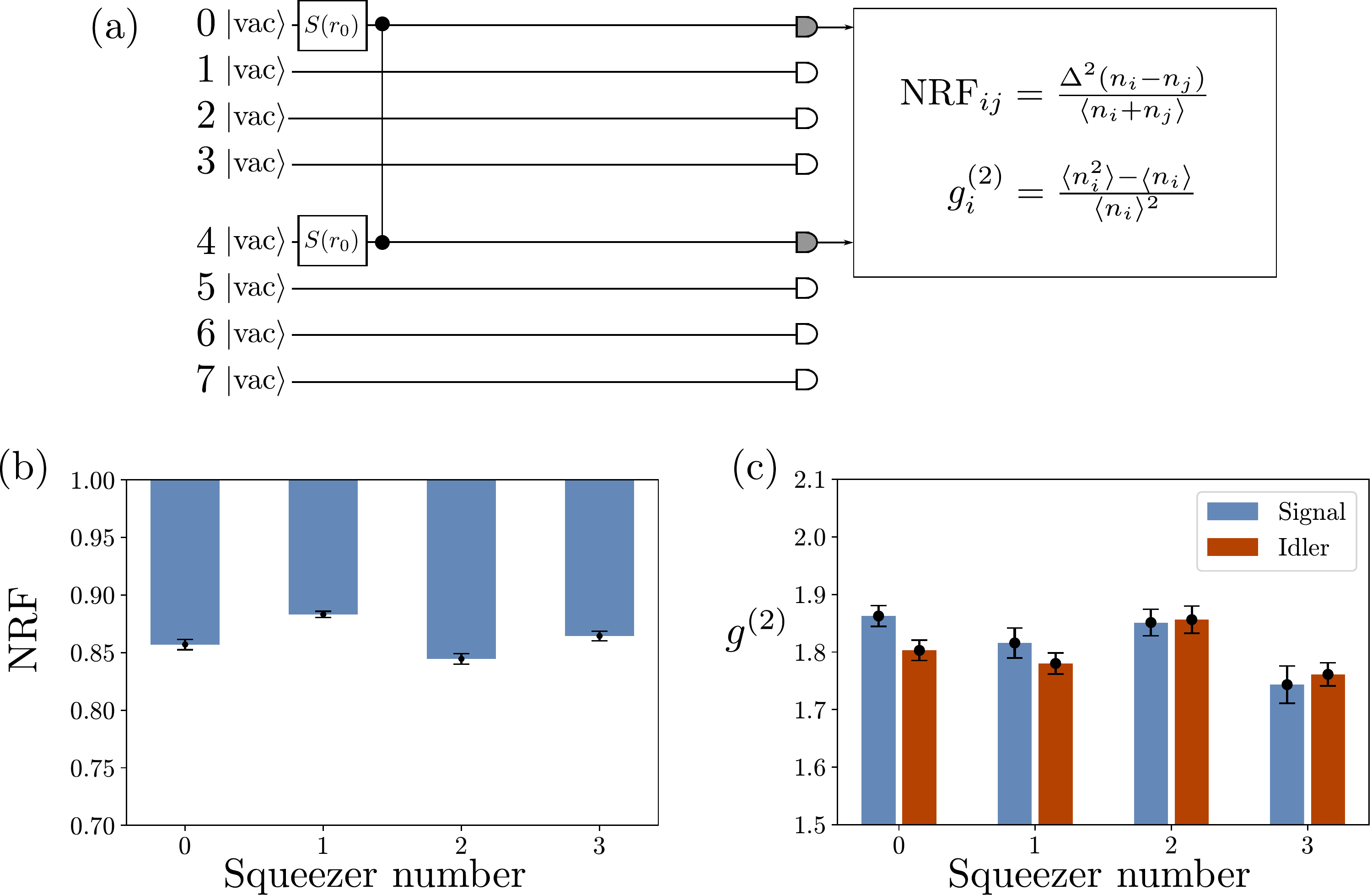}
    \caption{(a) Schematic of the circuit used to measure noise reduction factors and second-order correlation statistics for individual squeezers, here illustrated for squeezer $0$. The unitary is set to the identity transformation, and each squeezer is turned on individually. Photon samples collected from the corresponding signal and idler outputs are collected and used to calculate the relevant quantities. (b) Raw measured noise reduction factor (NRF) for each of the squeezers. Each is well below unity, indicating non-classicality. (c) Raw measured unheralded second-order correlation statistic $g^{(2)}$ of the signal and idler for each squeezer. Each is close to $g^{(2)}=2$, indicating nearly single temporal mode operation.}
    \label{fig:one-sq-stats}
\end{figure}

\begin{figure}[t!]
    \centering
    \includegraphics[width=0.9\columnwidth]{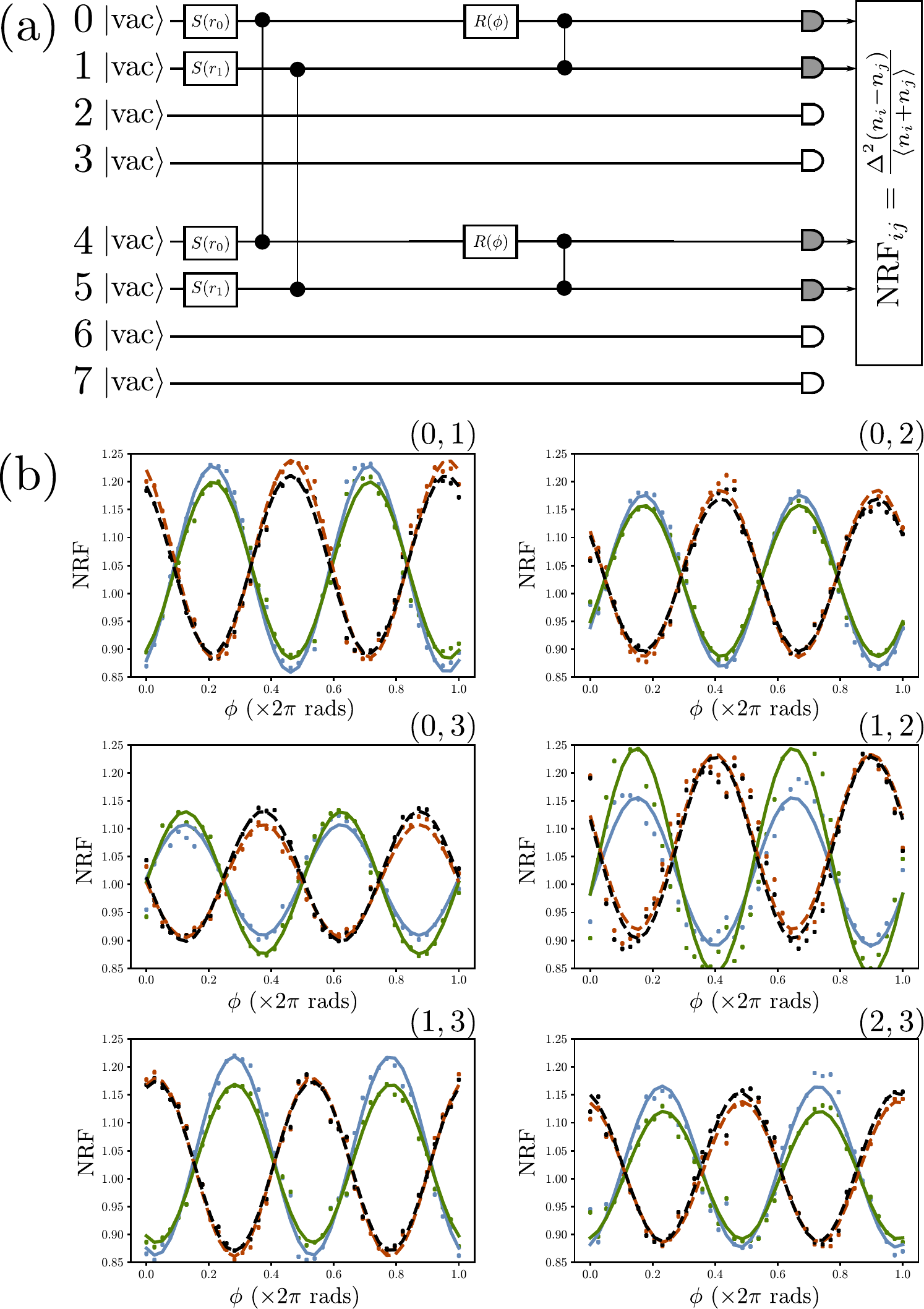}
    \caption{(a) Schematic of the circuit used to measure quantum interference between pairs of squeezers. Here the circuit for the $(0,1)$ pair is illustrated: two squeezers are turned on, and the interferometer is used to interfere their outputs on an effective 50/50 beam splitter with relative input phase $\phi$. The noise reduction factors are then calculated from the photon number samples. 
    (b) Interference traces between pairs of squeezers. The six panels each correspond to a different squeezer pair $(k,l)$. Within each panel, four noise reduction factors are plotted as function of the relative phase $\phi$: [signal 1 - idler 1] (blue), [signal 2 - idler 2] (green), [signal 1 - idler 2] (red), [signal 2 - idler 1] (black). Points correspond to raw, uncorrected measured data; solid and dashed lines are best fits (least squares) to a model that incorporates no imperfections except photon loss.}
    \label{fig:two-sq-interference}
\end{figure}

\subsection*{Noise reduction factor}
After calibrating the device, we characterize the component-level system performance by operating the interferometer in fixed simple configurations and computing relevant statistics on the photon counts acquired. As shown in Fig.~\ref{fig:one-sq-stats}(a), the interferometer is first set to the identity transformation and each squeezer individually turned on. The two-mode cross-correlation $V_{\Delta n}^{(i)}/n_{\mathrm{tot}}^{(i)}$ is then measured, where $n_\mathrm{tot}^{(i)}$ is the combined total mean photon number in the $i$th signal/idler mode pair and $V_{\Delta n}^{(i)}$ is the variance of the photon number difference between the $i$th signal/idler mode pair. This quantity is termed the noise reduction factor (NRF) and is a measure of non-classicality~\cite{aytur1990pulsed}. For two-mode Gaussian states, the NRF is a direct measure of the degree of entanglement between those two modes: $V_{\Delta n_i}/n_\mathrm{tot,i}=0$ indicates an ideal two-mode squeezed state, and $V_{\Delta n_i}/n_{\mathrm{tot},i}=1$ indicates a classical coherent state. As evident in Fig.~\ref{fig:one-sq-stats}(b), the measured NRF for each signal/idler mode pair is well below unity, averaging $0.86(1)$. This value is limited primarily by losses, which degrade the measurable correlations in an otherwise ideal two-mode squeezed state as $V_{\Delta n}^{(i)}/n_{\mathrm{tot}}^{(i)}=1-\eta_i$, with $\eta_i$ the total transmission efficiency experienced by mode pair $i$ (assuming balanced losses between the signal and idler pair). Our estimated system efficiency of approximately $15\%$, inferred both from direct measurements of components using classical light and from fitting the photon number statistics to a general theoretical model, is consistent with measured noise reduction factors. Based on this, we estimate the effective input squeezing in each mode (i.e., the squeezing produced by each squeezer in the circuit representation of Fig. 1, in the absence of losses) to be approximately 8 dB. This was chosen to correspond to a mean photon number of about one per mode at the sources, and could be increased by using more pump power or designing better resonators with higher escape efficiencies and quality factors. This value for effective input squeezing cannot easily be directly measured, but serves as a guideline for theoretical modelling of our device.

\subsection*{Second-order correlation}
For faithful execution of quantum circuits according to the idealized functionality illustrated in Fig.~\ref{fig:circuit_diagram}, it is important that no additional co-propagating modes are significantly populated with photons apart from those that carry the desired Gaussian state; since the photon detectors cannot distinguish between overlapping temporal modes, they would show up as an effective noise contribution to the collected samples. It is therefore vital to assess the temporal mode structure of the individual squeezer outputs: the squeezed states should as closely as possible populate only a single temporal mode. This can be quantified by the Schmidt numbers $K_i$~\cite{vaidya2019broadband,christ2011probing} of our squeezer sources, or, equivalently, the unheralded second-order correlation statistic $g^{(2)}_{S(I),i}=(\langle n_{S(I),i}^2\rangle - \langle n_{S(I),i}\rangle)/\langle n_{S(I),i}\rangle^2$, where $n_{S(I),i}$ is the photon number measured in the signal (idler) from the $i$th squeezer. This statistic is independent of the NRF of the sources, as it pertains not to the degree of photon-number correlation between the mode pairs, but to the temporal mode structure of each generated squeezed state. Ideally, $g^{(2)}_{S(I),i}=2$ for all squeezers, indicating a single-mode thermal state populating a single temporal mode, as is expected from each half of a two-mode squeezed state. The raw measured second-order correlation statistics for each of the eight measured modes is plotted in Fig.~\ref{fig:one-sq-stats}(c); the average $g^{(2)}$ over all eight modes is $1.81(4)$, indicating that our squeezers are working very close to single temporal mode operation. Based on this and the inferred level of background noise, we estimate that over $85\%$ of detected photons come from squeezing in the dominant Schmidt mode across all squeezers. {About $5\%$ arise from noise photons generated by Raman scattering in fiber components before the chip, and $10\%$ from unwanted temporal modes populated by the squeezers. These figure can be improved by implementing better wavelength filtering on the pump input to the chip to eliminate noise, and by engineering the squeezers to permit more broadband pump pulses to be used.}

\subsection*{Two-source interference}
An even more stringent requirement than single temporal mode operation is uniformity of the squeezed light sources: for high-visibility quantum interference to occur, the temporal modes populated by each squeezer must be nearly identical. To verify that genuine multi-source quantum interference is accessible in our device, we configure the interferometer to selectively interfere pairs of squeezed sources, and measure the phase-dependent response of four noise reduction factors between all six possible pairs of squeezers. A representative quantum circuit is shown in Fig.~\ref{fig:two-sq-interference}(a). The $24$ resultant traces are plotted in Fig.~\ref{fig:two-sq-interference}(b) alongside fits to a theoretical model of this interference that includes only optical loss as an imperfection. The pronounced phase-dependent response of the photon statistics, consistent with the theoretical model, demonstrates multi-photon quantum interference between all four sources. We emphasize that, in contrast to the typical presentation of data from experiments based on heralded single-photon sources, no post-selection or other post-processing was applied to the data exhibited in Fig. \ref{fig:two-sq-interference}(b).

The interference can be quantified by the amplitude of the oscillations in these traces. The NRFs between modes from separate squeezers, made to interfere according to the circuit of Fig. \ref{fig:two-sq-interference}(a), obey an oscillatory dependence on the relative phase $\phi$, with an amplitude proportional to the sum of the mean photon number (after losses) and total system transmissivity. The amplitudes extracted from the fits in Fig. \ref{fig:two-sq-interference}(b) are consistent to within $40\%$ of the independently estimated values for these quantities; imperfections apart from loss, including squeezer distinguishability, need to be accounted for in the model to obtain better agreement. This is described in more detail in the Methods section. In the future, more general modelling of the device can be used to extract an estimate for the overlap between the temporal modes populated by different squeezers, informing the path to optimizing two-source interference of these devices.
\subsection*{Non-classicality test}
Finally, we show that the output distribution of the device cannot be efficiently simulated with small error by approximating the output state with a classical Gaussian state, i.e., a state with positive Glauber-Sudarshan $P$-function~\cite{glauber1963coherent,sudarshan1963equivalence}. In other words, we show that the generated state has a non-positive $P$-function. This condition is necessary but not sufficient to demonstrate the inability to classically simulate the device.

To do this, we first characterize the chip by assuming a model with a single Schmidt mode per squeezer, non-uniform loss before the unitary transformation, and excess noise from residual photons not blocked by the filtering system, arising from the pump or broadband Raman scattering \cite{burenkov2017full}. The model parameters are determined by comparing its predictions with the actual samples collected. Using $P_0$ to denote the experimental photon number distribution and $P$ for the fitted model distribution, we find the sampling error, defined as $d_0:=\delta(P_0,P)$ where $\delta(P,Q)=\frac{1}{2}\|P-Q\|_1$ is the total variation distance, to be $d_0=0.10(1)$.

Now, a device described by the aforementioned noise model with high accuracy is deemed classical, meaning it can be efficiently simulated up to error $\epsilon$ by sampling from classical states, if the following condition is satisfied~\cite{qi2020regimes}: 
\begin{align}\label{eq:non-classicality-test}
    \sum_{i=1}^M \ln\pbrac{\frac{x_i+x_i^{-1}}{2}} < \epsilon^2/4~,
\end{align}
where $x_i = \sqrt{(\eta_ie^{-2r_i}+1-\eta_i)/(1-2p^{(D)}_i)}$, $\eta_i$ is the transmission efficiency of mode $i$, $r_i$ is the single-mode squeezing level, $p^{(D)}_i$ is the probability of detecting one excess photon, and $M$ is the number of modes. Setting $\epsilon$ equal to the modeling error $d_0$ and substituting the model parameters, we obtain $2.5\times 10^{-3}$ for the right-hand side and $1.0\times 10^{-2}$ for the left-hand side in Eq.~\eqref{eq:non-classicality-test}, meaning the inequality is not satisfied and the device passes the non-classicality test. The minimum error $\epsilon_0$ satisfying the inequality can also be interpreted as a measure of the minimal distance between the output noisy Gaussian state and all classical Gaussian states. We find that $\epsilon_0\approx 0.20$ for our device. This can be compared to previous four-mode experimental results~\cite{paesani2019generation} for which $\epsilon_0 \approx 0.017$ can be inferred~\cite{qi2020regimes}. Thus our device samples from a distribution that is quantifiably more non-classical than that of comparable devices \cite{paesani2019generation}. Physically, the higher degree of non-classicality of our device originates from the improved level of squeezing and transmission efficiency.

More details of the apparatus and component characterization procedure are available in the Methods section and Supplemental Information.

\section*{Demonstrations}

We showcase the programmability, high sampling rate, and photon number resolving capabilities of the machine by demonstrating proof-of-principle implementations of photonic quantum algorithms: Gaussian boson sampling (GBS), molecular vibronic spectra, and graph similarity. All three demonstrations use samples from the device to infer a property of the object central to the application. For GBS, the samples provide information about the non-classical probability distribution produced by the device. The vibronic spectra algorithm uses outputs from the device to obtain molecular properties, while for graph similarity, the samples reveal information on graph properties. In all demonstrations, the device is programmed remotely using the Strawberry Fields Python library \cite{killoran2019strawberry}, with example code shown in the Methods section. Theoretical predictions are performed with respect to a more detailed model of the device involving two Schmidt modes per squeezer, non-uniform loss before the unitary transformation, and excess noise. Model parameters are reported in the Supplemental Information.

\subsection*{Gaussian Boson Sampling}\label{sec:gbs}
Sampling from the distribution induced by a Fock basis measurement on Gaussian states is believed to require exponential time using classical computers~\cite{aaronson2013, hamilton2017gaussian}. This model is known as Gaussian boson sampling (GBS) and it is a leading platform being pursued for demonstrating a quantum advantage using photonic hardware. The complexity of the best known classical algorithms for simulating GBS scales exponentially with the number of detected photons, given a sufficiently large number of modes~\cite{quesada2020exact}. Therefore, generating a large number of photons is an important ingredient for achieving hardness of classical simulation.  There is also evidence that this remains the case provided losses remain sufficiently low~\cite{qi2020regimes}, but further work is needed to understand the experimental requirements for achieving quantum advantage with GBS. 

\begin{figure}
    \centering
    \includegraphics[width=0.85\columnwidth]{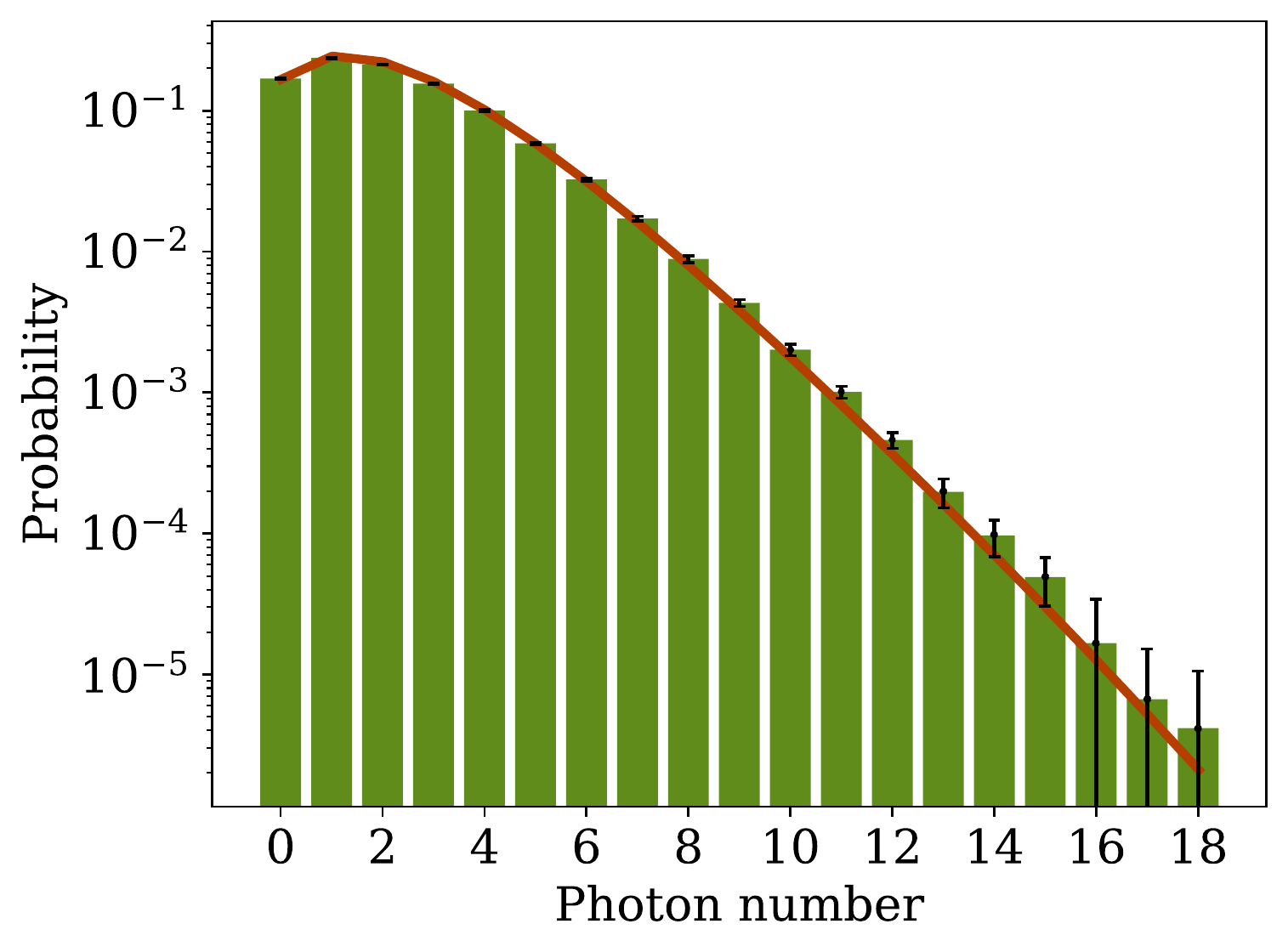}
    \caption{Probability distribution for the total number of photons generated by the device. All squeezers are turned on and the interferometer is set to the identity. Estimates of the probabilities obtained from experimental samples are shown as bars. The theoretical prediction appears as a continuous line. Error bars denote one standard deviation taken over 12 runs of $10^5$ samples. For large photon numbers, error bars are comparable to the probabilities.}
    \label{fig:photon_dbn}
\end{figure}

Due to strong on-chip squeezing in the device, a large number of photons can be generated. This is illustrated in Fig.~\ref{fig:photon_dbn}, which shows the probability distribution for the total number of photons measured. These values can be used to estimate the sampling rates for specific photon numbers relative to the raw sampling rate of $10^5$ events/s. 

\begin{figure}
    \centering
    \includegraphics[width=\columnwidth]{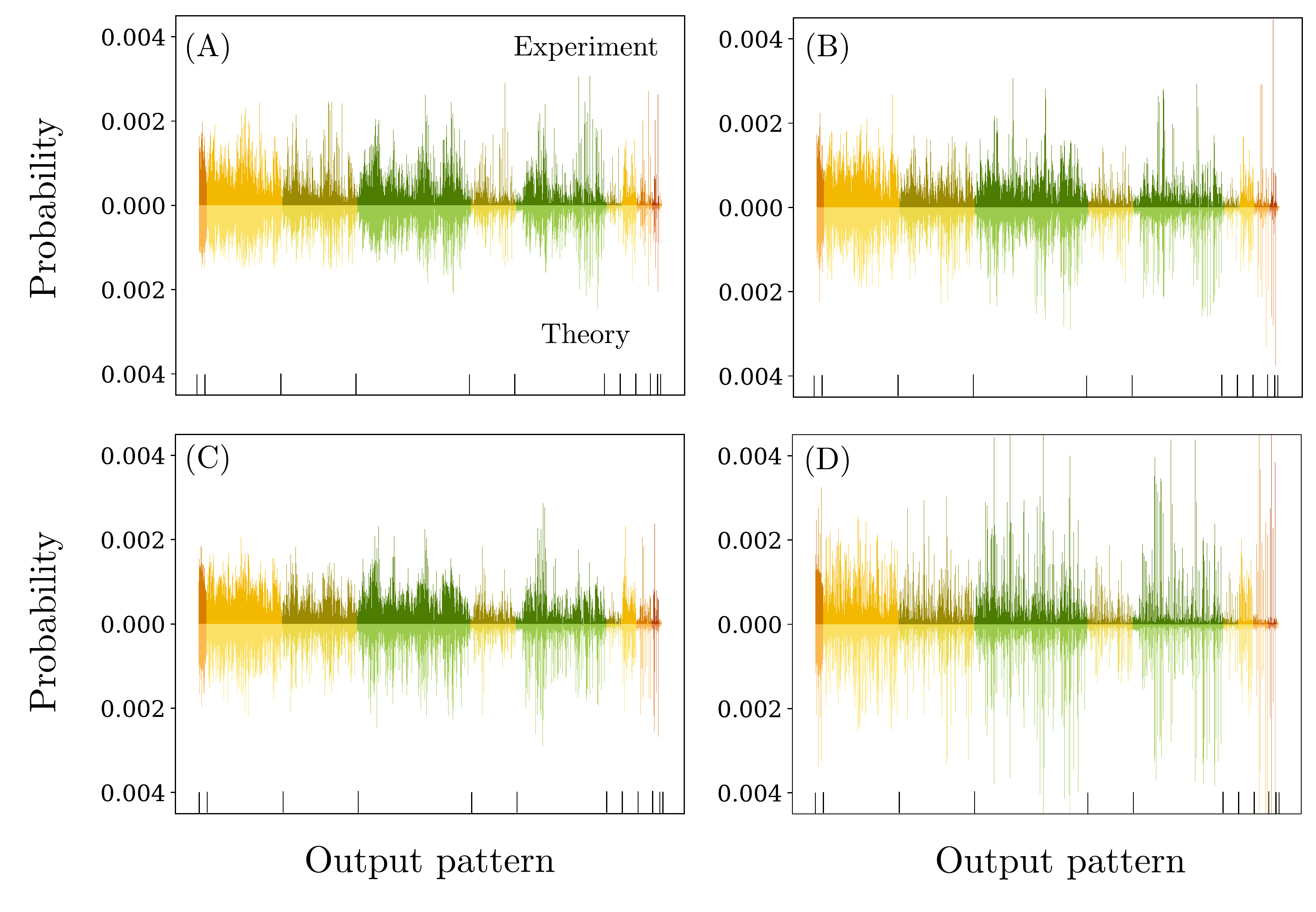}
    \caption{Probability distributions for six-photon outcomes in four separate GBS experiments. In each figure, the top bar plot depicts experimental probabilities estimated from chip samples and the bottom plots show the theoretical values. Output patterns are organized by orbits, separated by different colours as well as vertical bars in the bottom of the plots. Starting from the left,  the orbits are $[1,1,1,1,1,1]$, $[2,1,1,1,1]$, $[3,1,1,1]$, $[2,2,1,1]$, $[4,1,1]$, $[3,2,1]$, $[5,1]$, $[2,2,2]$, $[4, 2]$, $[3, 3]$, $[6]$. Plots $(A)$ to $(C)$ are the distributions for Haar-random interferometers and plot $(D)$ is the identity. }
    \label{fig:gbs}
\end{figure}

In the implementation, the device is configured according to three different interferometers randomly selected from the Haar measure, generating $1.2\times 10^6$ samples for each. For benchmarking purposes, sampling is repeated for an interferometer set to the identity. The results are shown in Fig.~\ref{fig:gbs}, where we plot the full distribution of six-photon output patterns compared to their theoretical predictions based on the detailed model described above. Outcomes are organized by permutationally-invariant patterns called orbits, which are sets of outputs determined by a specific photon detection pattern. As an example, the orbit $[3,1,1,1]$ corresponds to all six-photon outputs where there is one mode with three photons, three modes with a single photon, and zero photons in the rest. The average total variation distance between experimental and theoretical distributions, across all experiments, is $0.09(1)$. Previous state-of-the-art GBS experiments~\cite{zhong2019experimental} reported a maximum of five-photon events, whereas we observe 15-photon outputs with rates above 1 event/s. This makes our results the GBS demonstration with the largest number of detected photons to date. For reference, the largest boson sampling experiment reported 20-photon events across 60 modes, with a 14-photon coincidence rate of roughly 6 per hour~\cite{wang2019boson}. 

\subsection*{Vibronic spectra}\label{sec:vibronic}
The vibronic spectrum of a molecule specifies the frequencies and intensities of light absorbed when the molecule undergoes a transition between different vibrational and electronic states. Predicting vibronic spectra using classical methods is challenging because the Franck-Condon factors~\cite{sharp1964franck} that determine transition amplitudes generally require exponential time to be calculated. Nevertheless, photonic devices can be programmed to efficiently generate Franck-Condon profiles --- mathematical functions that determine the probability of observing a transition at a given frequency~\cite{huh2015boson}. In the photonic algorithm, optical modes represent the vibrational normal modes of the molecule and the device is programmed in terms of squeezing, displacement, and linear interferometers to efficiently generate Franck-Condon profiles~\cite{quesada2019franck,huh2015boson}. There is confidence that quantum algorithms for vibronic spectra can be scaled to outperform classical methods~\cite{sawaya2020near}, but work remains to support this convincingly.

\begin{figure}[t!]
    \centering
    \includegraphics[width=0.7\columnwidth]{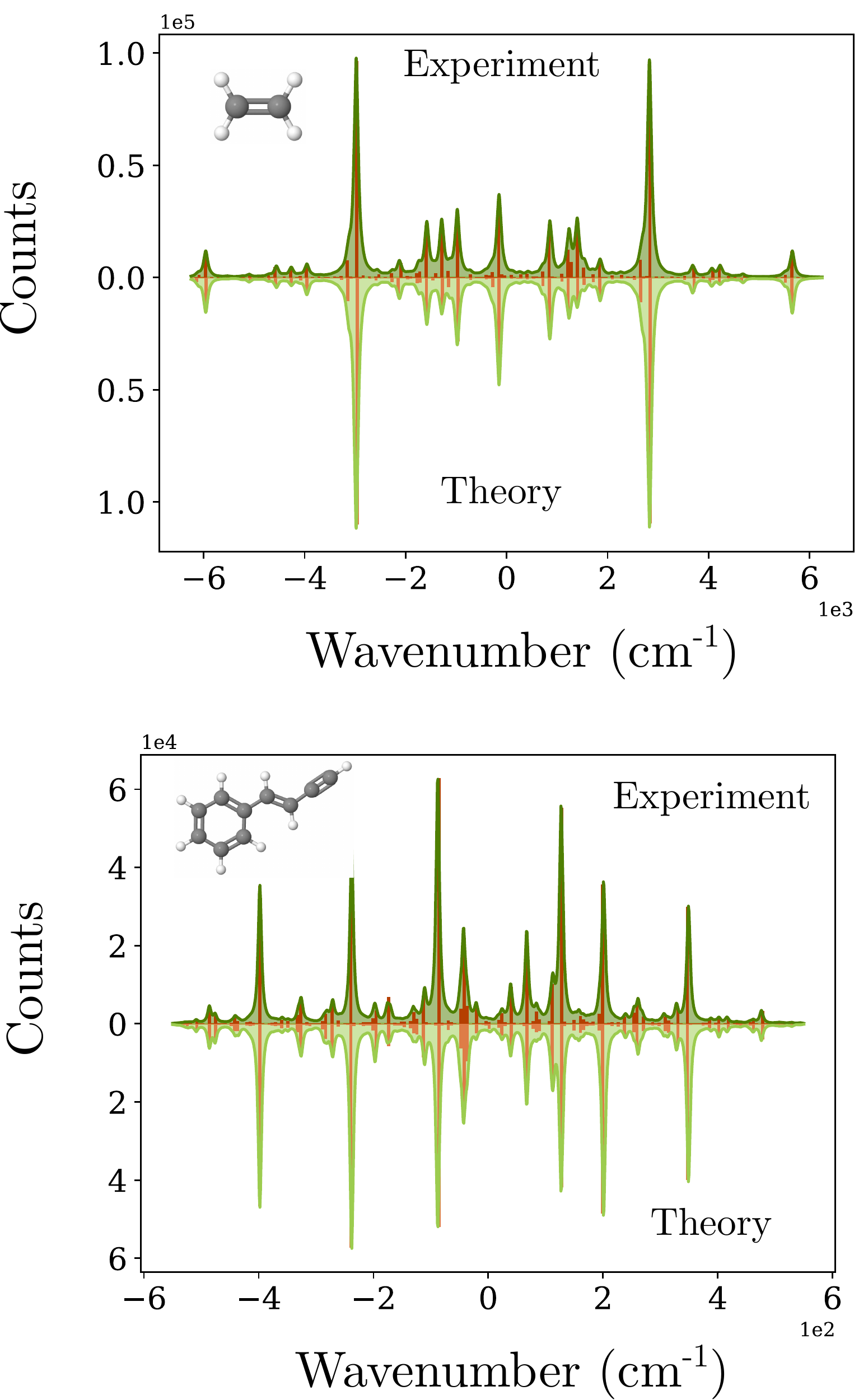}
    \caption{Franck-Condon profiles obtained from chip distributions programmed according to the vibronic transitions of ethylene (top) and (E)-phenylvinylacetylene (bottom). Red bar graphs depict the histogram of energies, while green continuous lines show a Lorentzian broadening of the bars. Wave numbers correspond to the energy differences between initial and final energy levels. Vacuum outputs are omitted: these correspond to zero-energy contributions resulting from transitions between vibrational ground states of the initial and final electronic states. There is in general strong agreement in the location and height of absorption lines between the experimental values and their theoretical model. }
    \label{fig:spectra}
\end{figure}

We program the chip interferometer according to the Duschinsky matrices that represent mixing between four normal coordinates in ethylene (\ce{C2H4}) and (E)-phenylvinylacetylene (\ce{C10H8}). Displacements are not included and squeezing is only present in the first mode, which means that the resulting profiles do not correspond to the true vibronic spectra of these molecules, which have displacements and different squeezing levels on each mode. Nevertheless they can be used as proof-of-principle benchmarks with respect to the theoretical model of the device: peaks in the reconstructed Franck-Condon profiles should coincide for both theory and experiment~\cite{paesani2019generation}. Results are shown in Fig.~\ref{fig:spectra}, obtained by generating $1.2\times 10^6$ samples for each molecule. Histogram bars are calculated using built-in functions from Strawberry Fields \cite{killoran2019strawberry} and are displayed for both molecules, together with a Lorentzian broadening of the bar, which is added to mimic the natural broadening that is observed in experiments. Peaks in the theoretical distribution, which correspond to absorption lines, are reproduced with similar intensity in the profiles reconstructed from chip samples, showing consistency between the theoretical model and the performance of the device. This indicates that, besides its applications to quantum chemistry, the vibronic spectra algorithm can be used as a benchmark for photonic devices.

\subsection*{Graph similarity}\label{sec:graph-sim}

Any undirected weighted graph can be encoded in a photonic circuit by exploiting a correspondence between real symmetric matrices --- representing the graph's adjacency matrix --- and the combination of a linear optical interferometer with squeezed light ~\cite{bradler2017gaussian}. The statistics of detected photon patterns contain information about the encoded graph, which can be used to quantify similarity between the graphs. One approach is to estimate orbit probabilities and collect them in $m$-tuples called feature vectors~\cite{schuld2020measuring,bradler2019duality}. The distance between the resulting feature vectors is used to quantify the similarity of the corresponding graphs. Initial studies support the merit of this algorithm~\cite{schuld2020measuring}, but it is currently unclear whether it can be scaled to outperform classical approaches, particularly in the presence of imperfections.

\begin{figure}[t!]
    \centering
    \includegraphics[width=0.95\columnwidth]{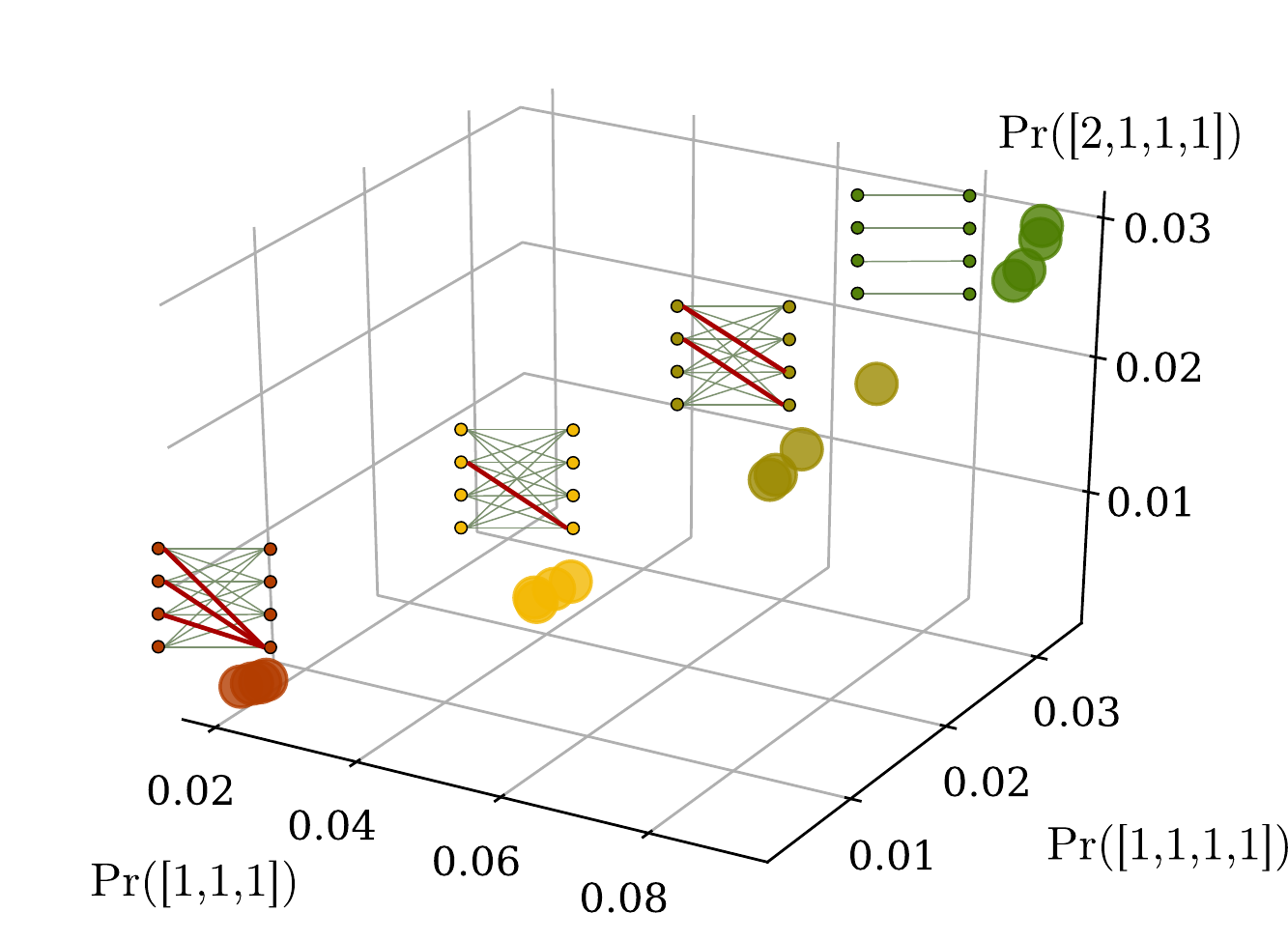}
    \caption{Feature vectors corresponding to four different graphs. The graphs are drawn next to their corresponding feature vectors, with negative-weighted edges highlighted by thick red lines. The components of the vectors are probabilities for the orbits $[1,1,1]$, $[1,1,1,1]$, and $[2,1,1,1]$, respectively. For each graph, feature vectors are also calculated for three random permutations of the graph. These appear as clusters of permutationally-invariant graphs, each cluster depicted by a different colour. }
    \label{fig:similarity}
\end{figure}

We demonstrate this algorithm by encoding bipartite graphs on eight vertices into the nanophotonic chip. Four graphs are considered, with their corresponding adjacency matrices shown in the Supplemental Information. Feature vectors are estimated with high statistical precision using 20 million samples for each graph. The results are illustrated in Fig.~\ref{fig:similarity}, showing that these graphs result in separate feature vectors. These feature vectors are built from orbits, which are permutationally-invariant sets of click patterns. It follows that isomorphic graphs have the same feature vectors. To showcase this property, three random permutations were selected and each of the four graphs was permuted accordingly. The results are again depicted in Fig.~\ref{fig:similarity}, resulting in clusters of isomorphic graphs, as expected from the permutation-invariance of the feature vector construction. These results are the first demonstration of graph similarity on a quantum device.

\vspace{-0.2cm}

\section*{Discussion}
The development and widespread deployment of quantum computing technologies is an ongoing worldwide effort, spanning several physical platforms. We have presented a nanophotonic device pioneering several record capabilities: high sampling rates, large on-chip squeezing, nearly ideal second-order correlation statistics, and significantly more detected photons than previously reported in similar devices. The hardware is programmable, and can be remotely configured via a custom application programming interface, using just a few lines of code. This software layer thus enables the deployment of our machine for cloud access. We have further showcased the capabilities of the nanophotonic chip with example demonstrations of Gaussian boson sampling, vibronic spectra, and graph similarity. The graph similarity demonstration is the first of its kind on any photonic platform, and the Gaussian boson sampling demonstration is the largest such experiment reported to date.

As the first of its generation, our device constitutes an initial step in scaling such nanophotonic chips to a larger number of modes. Doing so will enable reaching the regime of quantum advantage where classical simulation of the quantum device becomes intractable. The greatest challenge in scaling to a system of this size is maintaining acceptably low losses in the interferometer. New designs for integrated beamsplitters and phase shifters, requiring more precise (but readily available with current technology) chip fabrication tools, can achieve an order-of-magnitude improvement in the loss per layer in the interferometer. This would enable a 100-mode device in our architecture to be realized with less than $3$ dB of loss in the interferometer. More detail on the pathway to such improvements is available in the Scalability discussion of the Methods section. Furthermore, the inclusion of tunable single-mode (degenerate) squeezing \cite{vernon2019scalable} and displacement will constitute a significant upgrade, permitting the generation of arbitrary Gaussian states and unlocking the capability of implementing quantum algorithms with applications to quantum chemistry, graph theory, and optimization. Such scaling and upgrades are natural next steps for near-term photonic quantum information processing demonstrations. Our work represents a significant advance in the efforts to build a practical photonic quantum computer, and serves as a reference point for the rapidly progressing state of the art.

\appendix
\section*{Methods}

%\subsection*{Programming a nanophotonic chip}
%The device can be programmed remotely using the Strawberry Fields Python library \cite{killoran2019strawberry}. A user with valid credentials can specify the settings of the device using a few lines of code and subsequently request samples. The example Python code below, from version 0.14.0 of Strawberry Fields, shows a typical workflow where $4\times 10^5$ samples are requested from a device. All squeezers are turned on and the interferometer is programmed according to a unitary transformation drawn randomly from the Haar measure. 
%\begin{customcode}
%>>> import strawberryfields as sf
%>>> from strawberryfields import RemoteEngine, ops
%>>> from strawberryfields.utils import random_interferometer
%
%>>> U = random_interferometer(4) 
%>>> prog = sf.Program(8)
%>>> with prog.context as q:
%...    # Two-mode squeezing gates on all pairs of modes
%...    ops.S2gate(1.0) | (q[0], q[4])
%...    ops.S2gate(1.0) | (q[1], q[5])
%...    ops.S2gate(1.0) | (q[2], q[6])
%...    ops.S2gate(1.0) | (q[3], q[7])
%...    # Interferometer on the signal modes (0-3)
%...    ops.Interferometer(U) | q[0:4]
%...    # Interferometer on the idler modes (4-7)
%...    ops.Interferometer(U) | q[4:8]
%...    # Fock basis measurement
%...    ops.MeasureFock() | q  
%
%>>> # Run the program on the remote engine with 400k samples
%>>> eng = RemoteEngine('X8_01')
%>>> results = eng.run(prog, shots=400000)
%>>> samples = results.samples
%\end{customcode}

\subsection*{Apparatus details}
As described in the main text and in Fig.~\ref{fig:apparatus}, the full apparatus consists of:
\begin{itemize}
    \item A custom modulated pump laser source producing a regular pulse train ($100$ kHz repetition rate) of $1.5$ns duration rectangular pulses. 
    \item An electrically and optically packaged chip that synthesizes a programmable eight-mode Gaussian state with temporal mode characteristics appropriate for photon number resolving readout.
    \item A locking system which serves to align and stabilize the resonance wavelengths of the on-chip squeezer resonators. 
    \item An array of digital-to-analog converters (DACs) for programming phase shifter voltages on the chip.
    \item An array of low-loss (off-chip) wavelength filters to suppress unwanted light, passing only wavelengths close to the signal and idler for detection.
    \item A detection system, which consists of an array of eight transition-edge sensor (TES) detectors for photon number-resolving readout, and the auxiliary equipment required to operate and acquire data from them.
    \item A master controller consisting of a conventional server computer running custom software to coordinate the continuous and automated operation of all subsystems, and receive and process jobs sent to the machine.
\end{itemize}
In the following sections we provide more detail on these subsystems, and the techniques used to characterize them.

\subsection*{Pump system}
The pump laser is a compact continuous wave tunable laser assembly, tuned to a wavelength of $1554.9$ nm.  The laser is connected to a $10$GHz bandwidth fiber-integrated intensity modulator which is used to define a regular train of 1.5ns wide optical pulses with a $100$ kHz repetition rate.  The output of the modulator is coupled to a 99/1 fiber splitter, with the 1\% tap directed to a photodiode used to lock the modulator bias voltage.  Bias voltage locking in continuous operation is performed by a modular field programmable gate array (FPGA)/DAC board.  The other 99\% is directed to a fiber polarizer, before being sent to an erbium doped fiber amplifier (EDFA).  After the EDFA, the pump is spectrally filtered using low-loss fiber bandpass filters and directed to the chip subsystem.  All of the components of the pump are controlled remotely and do not require human intervention for operation.

\subsection*{Integrated components}
\begin{figure}
    \centering
    \includegraphics[width=\columnwidth]{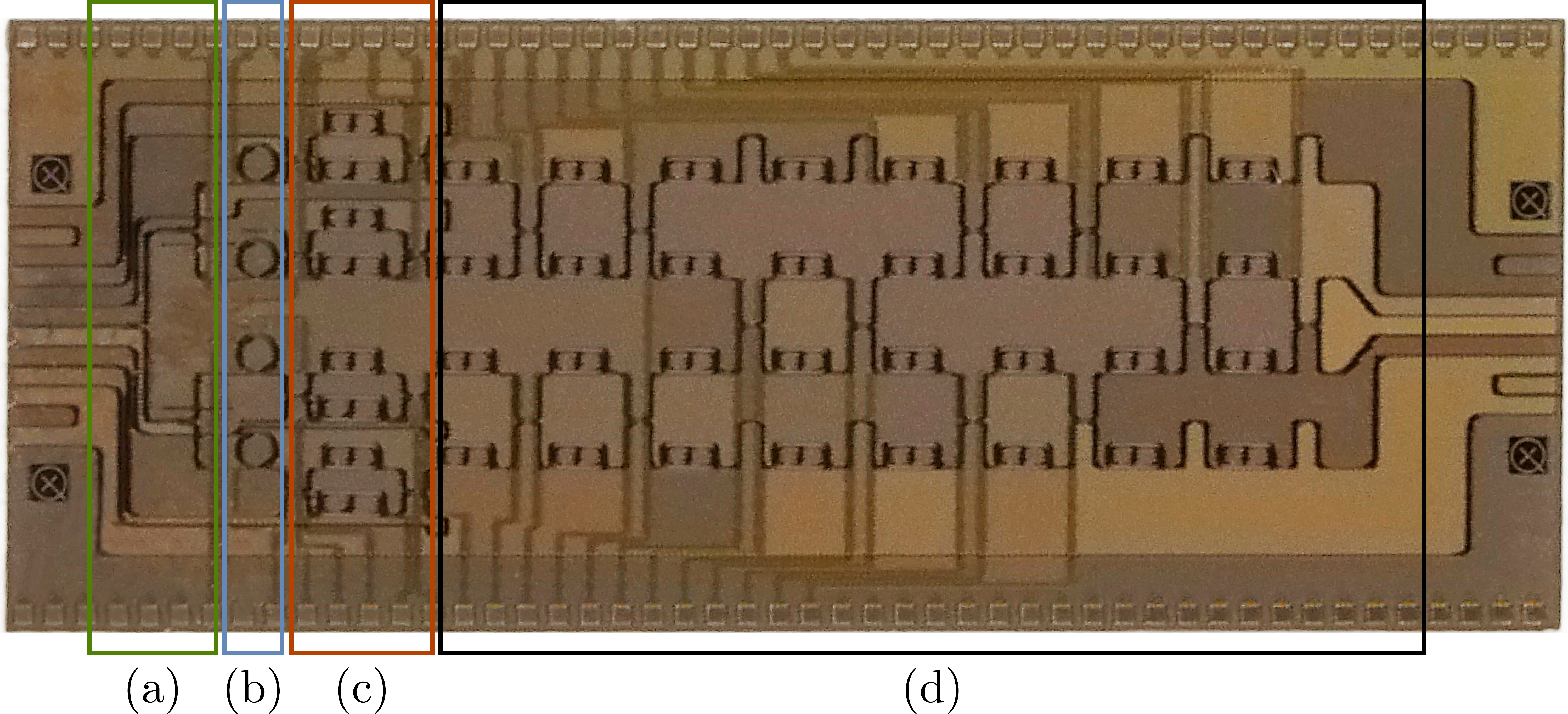}
    \caption{Micrograph of the full chip, showing the (a) input power distribution tree, (b) squeezer array, (c) AMZI filter array, and (d) programmable unitary transformation. The chip is approximately $10$ mm$\times 4$ mm in size.}
    \label{fig:chip_overview}
\end{figure}

The chip layout is shown in Fig.~\ref{fig:chip_overview}.  Pump light is edge-coupled from fiber to the chip through a single waveguide input.  This waveguide enters a binary tree of 50/50 beam splitters based on multimode interferometer (MMI) devices, which equally distributes the pump light among four spatial modes.  Each of these four waveguides is coupled to a separate squeezer. 

The squeezers are based on a microring resonator design that uses strongly pumped spontaneous four-wave mixing to generate bichromatic two-mode squeezing. This design is described in full detail by Vaidya \textit{et al.} \cite{vaidya2019broadband}; here we summarize the operation and details specific to the squeezers on the eight-mode chip. The waveguide cross-section of the rings is $1500$ nm x $800$ nm, and their radius is chosen to be $113$ $\upmu$m, corresponding to a free spectral range (FSR) of $200$ GHz. The loaded quality factors of the resonances used were approximately $7\times 10^5$, corresponding to a full-width-half-maximum linewidth of $275$ MHz, and varying less than 5\% across all four rings. The escape efficiencies for these resonances are approximately $75\%$, i.e., the probability of a photon generated in a ring being lost before it can be collected by the bus waveguide is approximately $25\%$. This makes up $1.2$ dB of the loss within the overall $8$ dB system efficiency.

To produce single temporal mode squeezed light, it is sufficient to employ pump pulses with duration comparable to the resonator dwelling time; the exact pulse shape is unimportant. In our case, $1.5$ ns square pulses yielded nearly-single temporal mode operation, as quantified by the second-order correlation data exhibited in Fig. \ref{fig:one-sq-stats}(c). Shorter pulses can be used, but do not appreciably improve the temporal mode structure, and compromise the generation efficiency as the pulse bandwidth exceeds the resonator linewidth. The exact pulse energy used is difficult to measure precisely, owing to the extremely low duty cycle of the pulse train, but we estimate this quantity to be on the order of 0.5 nJ.

No excess noise from unwanted processes occurring within the ring was measured. As discussed below, the dominant source of photon noise in the squeezing band is from Raman scattering in the fiber components carrying pump power to the chip. This can be managed in future versions by better pump filtering before the squeezers.

Each resonator output mode is directed to a separate asymmetric Mach-Zehnder interferometer (AMZI) device which acts as a pump rejection filter.  This ensures that no significant nonlinear light generation occurs in the interferometer portion of the chip, and also allows the rejected pump to be collected and used as a signal for locking the ring resonances to the pump laser wavelength. The bright outputs of the AMZI filters are directed back to the input facet and coupled out of the chip for detection. The FSRs of the AMZIs and rings are carefully matched to be compatible with the standard telecom dense wavelength division multiplexing (DWDM) spacing of 100GHz, and to allow the signal and idler to pass to the interferometer when the AMZI is tuned to reject the pump. The signal and idler resonances are each separated in frequency from the pump by three ring FSRs (approximately $600$GHz).

The interferometer is composed of a network of MMIs and phase shifters in a rectangular configuration~\cite{clements2016optimal}. This configuration contains a sequence of six $SU(2)$ transformations that enable arbitrary programmability of the interferometer by controlling the thermo-optic phase shifters integrated within the chip. The splitting ratio of the MMIs is constant to within $1\%$ over the range of wavelengths used. This control is accomplished using a multi-channel DAC system.  Light is coupled out of the chip via edge couplers to a fiber array, and then directed to a fiber-based low-loss filter stack that separates the signal and idler photons and directs them to separate photon number resolving detectors. The total pump rejection ratio is well in excess of $100$ dB. In addition, the filter stack rejects photons from unwanted resonator modes, and any residual pump light and broadband generated photons from in-fiber Raman scattering. The total remaining number of noise photons per pulse from all sources (pump leakage and Raman scattering) incident on the TES detectors is approximately $0.02$ or lower for each channel. The residual pump light rejected by the filter stack is directed to a photodiode array, and was used for the calibration of the interferometer. The filter stack comprises approximately $2$ dB of the overall $8$ dB of loss in the system.

The chip is both electrically and optically packaged to ensure stable operation. The chip is glued to a copper sub-mount using a thermally conductive die adhesive. The submount is mounted on top of a thermo-electric cooler used to actively stabilize the temperature of the chip.  Connectorized printed circuit boards (PCBs) are affixed to the sub-mount and the chip is wirebonded to these boards. Cables carry the electronic signals responsible for programming the unitary transformation and locking the rings to a secondary PCB which interfaces with custom control circuitry and the interferometer DAC. V-groove arrays of ultra-high numerical aperture (UHNA7) fiber are aligned to each edge facet of the chip using loop-back waveguide structures placed on the chip. These fiber arrays are affixed in place using an optical adhesive, resulting in an average coupling efficiency of approximately $70\%$.

\subsection*{Operating procedure}
Quantum programs are written by users with the Strawberry Fields Python library \cite{killoran2019strawberry}. These programs are sent to the master controller as ``jobs", i.e., scripts specifying squeezing parameters and interferometer phases. Upon receipt of a job, the information is compiled into a set of hardware instructions.  The control system then implements the following control sequence:
\begin{itemize}
    \item Voltages of the interferometer that correspond to the requested unitary operation are set.
    \item The chip is allowed to thermally equilibrate.
    \item The ring resonance wavelengths are swept to calibrate the squeezer control circuitry, followed by locking of the rings to the pump wavelength.
    \item Checks are performed to ensure that the interferometer and squeezers are in the desired state.
    \item The requested number of samples are acquired from the detectors.
    \item Checks are performed to ensure the interferometer and squeezers are still in their desired state, i.e., that the chip has not drifted out of the specified state during data acquisition.
    \item The sample and job data is returned to the user.
    \item The chip is re-initialized to its default state.
\end{itemize}

\subsection*{Chip calibration}
In order to set the interferometer to a user-specified state, the on-chip thermo-optic phase shifters must first be calibrated to determine the voltage-to-phase relationship for each phase shifter. The thermal nature of the phase shifter implies (and tests confirm) that to a high degree of accuracy, the relationship between phase and voltage can be described by:
\begin{equation}
\phi = \phi_0 + \alpha V^2    
\end{equation}
The goal of the calibration process is to determine $\phi_0$ and $\alpha$.  Then, when a specific phase is requested, the phase-to-voltage can be inverted to produce the required voltage.  The calibration is accomplished by injecting classical light into a single mode of the interferometer at a time by injecting pump light into the second input of the filter AMZI for that mode. A standard telecom fiber switch enables control of which mode the calibration light is injected into. The transmission of the interferometer is detected using classical light detectors connected to the pump rejection channel of the output filter stack. Employing optimization algorithms, it is possible to learn the voltage-to-phase relationship for each thermo-optic phase shifter in sequence.  

It is challenging, however, to learn the input phases of the interferometer using classical light, since these phases will depend on properties of the squeezers themselves.  Instead, to calibrate these three relevant phases, two-squeezer interference is used. Each pair of neighbouring squeezers is locked to the pump laser, and the input phase shifters in modes 0, 1, and 2 are swept. Mode 3 has no input phase shifter since only the relative phase between the inputs is physically relevant. The NRF is monitored between the pair of interfering modes and the relevant phase-to-voltage relationship is extracted. 

\subsection*{Photon detection system}
Each of our TES-based detectors has quantum efficiency above $95$\% and produces an analog voltage pulse every $10$ \si{\micro\second}, synchronized with the incident optical pulse train, with a shape that depends on the number of incident photons. These voltage signals are digitized by analog-to-digital converters, resulting in time series referred to here as voltage traces. Thus, determining photon numbers amounts to being able to associate a photon number $n$ to each trace. This is typically accomplished for sets of a few hundred thousand traces, by first ordering them according to a feature like their maximum or their overlaps with some reference trace. Reasonable points are then  determined, in terms of this feature, by which to organize the traces into photon-number bins~\cite{levine2012algorithm,humphreys2015tomography}. In previous work on measuring photon number difference squeezing from nanophotonic sources~\cite{vaidya2019broadband}, a principal component analysis was performed on sets of $8\times 10^5$ traces. These traces were then ordered with respect to their overlap with their first principal component, and a sum of Gaussians fit to the resulting histogram, solving for the points of intersection between adjacent Gaussians to determine photon-number bin edges. 

That approach suffers from two drawbacks which make it less appropriate for a more complex system like that described in this work. The first is that it relies on a global comparison of each trace to the full set of traces acquired during the corresponding experimental run, and so cannot associate a photon number with a single trace in real time after it is generated given that the principal component analysis depends on all traces in the data set. This limits the speed of the trace-to-photon number discrimination in our system. Second, and of more concern, the maximum assignable photon number $n_{\text{max}}$, i.e., the $n$ at which actual $(n+m)$-photon events (with $m>0$) will be identified as $n$-photon events, could be different for each data set, as each data set may identify a different number of photon-number bins. Both these drawbacks were eliminated in our system.

Before activating the full system, we first calibrate each detector, allowing each subsequent voltage trace to immediately be assigned, in real time, to a photon number up to the $n_{\text{max}}$ determined by the calibration. This calibration involves two steps: (i) identification of a standard trace for calculating overlaps, and (ii) determination of photon-number bin edges associated with the standard trace. Each calibration uses a set of $10^7$ voltage traces. To obtain a standard trace, we perform principal component analysis and histogram fitting to identify all of the two-photon traces in the set, and calculate the resultant average trace.  We use the set of two-photon traces as opposed to one-, three-, or four- photon traces in an effort to balance the tradeoff between capturing some detector nonlinearity and having enough events to obtain a representative average trace. Using sets of higher photon-number traces in principle allows us to extend $n_{\text{max}}$. However, as we calibrate using one arm of a two-mode squeezed vacuum state we always expect to have more $n$- than $(n+1)$-photon traces. Next, we calculate the overlap of each trace in the full set of $10^7$ traces with the standard trace, generate a histogram, fit to it a sum of Gaussians, and determine photon-number bin edges. The resultant $n_{\text{max}}$ for each of our eight detectors ranges between five and seven.

\subsection*{Noise reduction factor}
To assess the degree of photon number correlations between the signal and idler for each individual squeezer, the noise reduction factor (NRF) was measured. For a single two-mode squeezed vacuum source, we define this as
\begin{equation}
    \mathrm{NRF}=\frac{\Delta^2(n_s - n_i)}{\langle n_s + n_i \rangle},
\end{equation}
where $n_s$ and $n_i$ are the photon number observables for the signal and idler, respectively, and $\Delta^2(n_s - n_i)$ refers to the variance of the photon number difference. An ideal measurement of a perfect source would yield $\mathrm{NRF}=0$, since the photon number of the signal and idler are perfectly correlated for a two-mode squeezed vacuum state. On the other hand, a pair of coherent states would yield $\mathrm{NRF}=1$. In our system, the dominant imperfection that degrades the correlation is loss: a total photon transmission efficiency of $\eta$ yields an NRF of
\begin{equation}
    \mathrm{NRF}=1 - \eta
\end{equation}
for two-mode squeezed vacuum~\cite{vaidya2019broadband}.

The NRF values reported in Fig.~\ref{fig:one-sq-stats}(a) of the main text were obtained by setting the interferometer to the identity transformation, activating only one squeezer at a time, and collecting $8\times 10^5$ samples. These samples were divided into eight batches of $1\times 10^5$, and the NRF calculated for each batch. The mean and standard deviation of these eight NRF values  correspond respectively to the data points and uncertainties ($\pm 1\sigma$) reported. 

\subsection*{Second-order correlation}
To verify that each squeezer is significantly populating only one temporal mode, the unheralded second-order correlation statistic $g^{(2)}$ was measured for the signal and idler of each squeezer~\cite{vaidya2019broadband}. For any output channel of the device described by photon number operator $n$, this statistic is defined as
\begin{equation}
    g^{(2)} = \frac{\langle n^2 \rangle - \langle n \rangle}{\langle n \rangle^2}.
\end{equation}
This statistic provides a loss-insensitive measure of the temporal mode structure of a two-mode squeezed vacuum source. In the absence of noise, the Schmidt number $K$ is related to $g^{(2)}$ via~\cite{christ2011probing}
\begin{equation}
    g^{(2)}=1+\frac{1}{K}.
\end{equation}
An ideal single-temporal-mode two-mode squeezed vacuum source would yield $g^{(2)}=2$ for the signal and idler, whereas coherent states or highly multi-mode squeezed light would yield $g^{(2)}=1$.

The $g^{(2)}$ values reported in Fig.~\ref{fig:one-sq-stats}(b) were obtained, like the NRF values, by setting the interferometer to the identity transformation, activating only one squeezer at a time, and collecting $8\times 10^5$ samples. These samples were divided into eight batches of $1\times 10^5$, and the $g^{(2)}$ calculated for each batch. The mean and standard deviation of these eight $g^{(2)}$ values correspond respectively to the data points and uncertainties ($\pm 1\sigma$) reported. The values reported are raw and uncorrected for noise, which would tend to lower the measured $g^{(2)}$ towards unity. Noise from unwanted Raman scattering is the dominant factor affecting the measured $g^{(2)}$ in our system, and therefore the values reported are in fact lower bounds for this quantity.

\subsection*{Two-squeezer interference}
Here we provide a simple model to explain the behaviour of the noise reduction factor as a function of the phases of the interferometer used in our chip.  We consider two identical squeezing sources, labelled 1 and 2, that each produce photons in their idler arms $a_1$, $a_2$ and in their signal arms $b_1$ and $b_2$. We write the noise reduction factor between an arbitrary pair of modes $c,d$  as
\begin{eqnarray}
\text{NRF}_{cd} &=& \frac{\Delta^2 (n_c-n_d)}{\braket{n_c+n_d}}  \\
&=& \frac{ \Delta^2 n_c + \Delta^2 n_d - 2 \left( \braket{n_c n_d} - \braket{n_c} \braket{n_d} \right)}{\braket{n_c+n_d}}. \nonumber
\end{eqnarray}
Since we are considering Gaussian states (two-mode squeezed states with squeezing parameter $r$) undergoing Gaussian operations (a beam splitter with unitary matrix $U$ and loss quantified by transmission efficiency $\eta$), and assuming the losses to be homogeneous and the squeezing identical in both sources, it can be shown that the variance and mean photon number of all the modes are the same and given by
\begin{align}
\Delta^2 n = \bar{n} (\bar{n}+1), \quad 
\braket{n}  = \bar{n} = \eta \sinh^2(r).
\end{align}
Now we only need to evaluate
\begin{align}
\braket{n_c n_d} = \braket{c^\dagger c d^\dagger d} = \braket{c^\dagger c} \braket{d^\dagger d} + \braket{c^\dagger d^\dagger} \braket{c d},
\end{align}
where Wick's theorem \cite{vignat2012generalized} was used to write the fourth order expectation values in terms of second order ones.
For our system, the same interferometer acts on both the signal modes and the idler modes (per Fig.~\ref{fig:circuit_diagram} in the main text), and that interferometer transformation can be expressed according to $a_i \to \sum_{j} U_{ji} a_j$. With this, we find that
\begin{eqnarray}\label{NRFs}
\text{NRF}_{a_1,b_1} &=& \text{NRF}_{a_2,b_2} \\ 
&=&1-\eta+ (\eta +n)  \sin ^2( \theta ) \sin ^2(\phi ), \nonumber \\
\text{NRF}_{a_1,b_2} &=& \text{NRF}_{a_2,b_1} \nonumber \\
&=&1+n-(\eta+n)  \sin ^2(\theta ) \sin ^2(\phi ),\nonumber
\end{eqnarray}
where we parametrized the interferometer in terms of the unitary matrix
\begin{align}
U = \left( \begin{matrix}
\cos \theta/2 & e^{i \phi}\sin \theta/2 \\
-e^{-i \phi}\sin \theta/2 & \cos \theta/2
\end{matrix} \right).
\end{align}

The data exhibited in each panel of Fig.~\ref{fig:two-sq-interference}(b) were obtained as follows: The corresponding pair $(k,l)$ of squeezers were activated, with the others turned off. The unitary transformation $U$ was set to interfere the two squeezers with $\theta=\pi/2$, corresponding to an effective 50/50 beam splitter with relative input phase $\phi$. A batch of $4\times 10^5$ photon number samples was then acquired for each of $40$ different settings of $\phi$ between $0$ and $2\pi$. The four NRF combinations (signal 1-idler 1, signal 2-idler 2, signal 1-idler 2, signal 2-idler 1) were then computed from these samples, and the results plotted alongside least-squares fits to the model of Eq. (\ref{NRFs}) (with a free offset phase included to account for calibration offsets in $\phi$). 

Note that if the sources were completely distinguishable, i.e., if the temporal modes populated by different squeezers were very different, then the visibility of the interference would be zero: interferometer would not be able to interfere the modes and there would be no oscillating phase dependence with amplitude $n+\eta$ in Eq.~\eqref{NRFs}. This can be quantified partially by the extracted fit parameters for the curves, which when averaged over all traces give $n=0.18(4)$ and $\eta=0.11(1)$. The extracted transmission efficiency is consistent with independent estimates, whereas the extracted mean photon number is about $40\%$ smaller than independent estimates. The interference visibility is thus measurably affected by imperfections other than loss, including unitary transformation infidelity (the effective 50/50 beam splitter has approximately $18$dB extinction), noise, temporal multi-modedness, and potentially some squeezer distinguishability. 

\subsection*{Scalability}
An important factor in assessing the viability of the platform presented is the scalability of this approach. What improvements to the platform and design are required in order to scale the system size to a level where quantum advantage could potentially be achieved? To answer this, we fix a target of 100 modes, which in our architecture would require: 50 squeezers operating with squeezing factors of $r\approx 1$, a universal 50-spatial-mode interferometer, and 100 PNR detector channels. We also stipulate, as a rough estimate, that such a machine should incur no more than $3$ dB of loss in the interferometer; this criterion is especially demanding, since the interferometer loss scales with the number of modes. Events with hundreds of photons would be detectable with such a machine.

Presently, the total system loss is approximately $8$ dB, of which about $3$ dB is incurred in the four-spatial-mode interferometer. This is dominated by losses in the MMI-based beam splitters ($0.2$ to $0.4$ dB per layer) and in the bent segments of the waveguide coils used in the interferometer phase shifters ($0.35$ to $0.55$ dB per layer). MMIs are employed for their fabrication tolerance, as they reliably achieve close to 50:50 splitting ratio across large chip areas even with imperfect lithography and wafer uniformity. The waveguide coils are designed to achieve a longer phase shifter propagation length, increasing thermal efficiency. For both of these components, the dominant source of loss is not directly related to the fundamental straight-waveguide propagation loss of $0.2$ dB/cm associated with their lengths. 

Optimization of the design and fabrication process can greatly reduce these losses. By moving to a fabrication line offering more precise lithography, less fabrication-tolerant directional couplers can replace MMIs as the beam splitting element. These can achieve length-limited loss, contributing approximately $200 \upmu$m length per layer, which would correspond to about $0.008$ dB of loss per layer. Upgrading the microheaters used in the phase shifters to a more specialized material can lower the required number of bends and shorten the propagation length of the two waveguide coils to $3$ mm per layer, contributing $0.06$ dB per layer. These coils can also achieve length-limited performance by designing more adiabatic transitions between straight and bent segments. Combined, these changes would yield an interferometer loss of approximately $0.068$ dB per layer. For a 50-spatial-mode interferometer, this would result in a total of $3.4$ dB of loss. A modest improvement in waveguide propagation loss to $0.17$ dB/cm would then suppress interferometer losses to below $3$ dB. Considering silicon nitride waveguides have been demonstrated in a similar platform with losses as low as $0.055$ dB/cm \cite{pfeiffer2018photonic}, we believe this is a demanding but realistic pathway to controlling losses as the system size scales.

Other challenges associated with scaling the interferometer arise from the power dissipated by the thermo-optic phase shifters. Currently, the interferometer in our device dissipates approximately $1$ W of power for a typical unitary setting, in a chip area of $0.4$ cm$^2$. A 50-spatial-mode interferometer would require 2,450 phase shifters, dissipating a total of about $120$ W across a chip area of about $21$ cm$^2$ (corresponding to three reticle write-fields of a standard lithography tool), when each is tuned to achieve a $\pi$ phase shift. The thermal load density (power dissipated per unit chip area) would therefore approximately double, despite the number of phase shifters increasing by two orders of magnitude. For comparison, a modern microprocessor dissipates between $100$ and $200$ W under full load in a die area of about $1$ cm$^2$. With proper thermal management, we do not anticipate power dissipation posing a significant barrier to scaling.

\vspace{0.1in}
\emph{Certain commercial equipment, instruments, or materials are identified in this paper to foster understanding. Such identification does not imply recommendation or endorsement by the National Institute of Standards and Technology, nor does it imply that the materials or equipment identified are necessarily the best available for the purpose.}

\emph{All data and code required to evaluate the conclusions of this work are available from the authors upon reqeust.}
\bibliographystyle{ieeetr}
\bibliography{references}

\onecolumn

\begin{center}\section*{Supplemental Information}\end{center}
\subsection*{Model parameters}
A theoretical model of the chip distribution is used for benchmarking purposes in the experimental demonstrations. To estimate the model parameters quoted in the tables below, we construct a two-dimensional photon-number histogram for each signal and idler mode in a two-mode squeezed vacuum state generated by a single squeezer, keeping all other squeezers off. We model this data as a pair of two-mode squeezed vacua (two Schmidt modes each with squeezing parameter $r_i$) hitting the detectors after undergoing loss (with transmissivity $\eta$). The squeezing parameter is related to the two-mode squeezing operator by $S_2(r) = \exp\left(r ( a^\dagger b^\dagger - ab ) \right)$.
To represent noise in the detectors, we add an extra model with Poisson statistics (mean value $\bar{n}$) that accounts for the measured counts when all the squeezers are off. With these physical parameters it is possible to calculate a two-dimensional histogram using the methods from Ref.~\cite{burenkov2017full}. After this we simply use the well-known Levenberg–Marquardt algorithm to solve the inverse problem and retrieve the physical parameters from the measured photon number histograms. It is important to note that these parameters are not the directly measured values of squeezing and losses; they are the values that best approximate the behaviour of the chip given the simplified model we consider.

\begin{table}[h]
\centering
\begin{tabular}{|c||c|c|c|c|c|c|c|c|}
\hline
    Mode & 0 & 1 & 2 & 3 & 4 & 5 & 6 & 7 \\
    \hline 
    Transmissivity $\eta$ & 0.154(2) & 0.120(2) & 0.173(2) & 0.139(2) & 0.152(2) & 0.128(2) & 0.197(2) & 0.170(2)\\
    Noise $\bar{n}$ & 0.0205(1) & 0.0091(1) & 0.0131(1) & 0.0193(1) & 0.0223(1) & 0.0175(1) & 0.0158(1) & 0.0187(1)\\
    \hline
\end{tabular}
\caption{Transmissivity and background noise parameters for the theoretical model of the device. The noise values denote the mean photon numbers measured when all squeezers are off. }
\end{table}

\begin{table}[h]
\centering
\begin{tabular}{|c||c|c|c|c|}
\hline
    Modes & (0, 4) & (1, 5) & (2, 6) & (3, 7)  \\
    \hline 
    Squeezing $(r_1)$ & 1.162(6) & 1.101(7) & 1.050(5) & 1.005(6)\\
    Squeezing $(r_2)$ & 0.345(9) & 0.367(9) & 0.393(7) & 0.336(9)\\
    \hline
    
\end{tabular}
\caption{Two-mode squeezing parameters for the theoretical model of the device. The parameter $r_1$ denotes the squeezing level of the main Schmidt mode, while $r_2$ is the parameter for the second Schmidt mode.}
\end{table}

\hspace{1cm}

\subsection*{Sampling from non-classical light}
A non-classicality test for photonic devices has been formulated by Qi \emph{et al.} \cite{qi2020regimes}. The results there presented are valid for a simple noise model that includes uniform single Schmidt mode squeezers, uniform loss, and threshold detectors with dark counts. Therefore, we also consider a model with a single Schmidt mode and coarse-grain the output distribution as if obtained with threshold detectors. We furthermore generalize the formula in~\cite{qi2020regimes} to include non-uniform squeezing and losses. Numerically, we find a modeling error of $d_0=0.10(1)$ averaged over 15 random unitary transformations and calculations are made by considering a cutoff of 14 photons per mode. Since the coarse-graining procedure can only decrease the total variation distance, we can use the value of $d_0$ quoted above. 

We briefly present the derivation of Eq.~\eqref{eq:non-classicality-test} in the main text, which generalizes the results of Ref.~\cite{qi2020regimes}.  
Assuming the aforementioned noise model, the output quantum state of the device is given by $\rho = U (\prod_{i=1}^M\sigma_i) U^\dagger$, where $\sigma_i = L_{\eta_i}(\ket{r_i}\bra{r_i})$ are the lossy squeezed states in each mode. In Ref.~\cite{rahimi2016sufficient}, the authors studied the problem of exact sampling from an $M$-mode quantum state of the form $\tilde{\rho}= U(\prod_{i=1}^M \tau_i)U^\dagger$, where $\tau_i$ is an arbitrary $(t_i)$-classical Gaussian state, i.e., a state with positive $s_i$-ordered phase-space quasiprobability distribution~\cite{rahimi2016sufficient}. We denote the distribution obtained by sampling from this classical state by $\tilde{P}$, which is calculated using The Walrus~\cite{gupt2019walrus}. It can be shown that sampling from $\tilde{\rho}$ by using noisy threshold detector with excess photon rate $p_i^D$ can be simulated exactly in classical polynomial time if $t_i>1-2p_i^D$~\cite{rahimi2016sufficient}.

Therefore, when the mixed input state $\sigma_i$ is close to some classical Gaussian state $\tau_i$, the corresponding noisy GBS experiment can be efficiently simulated with small error. Since any such state $\tau_i$ leads to an efficient classical simulation, it is necessary to minimize the distance to $\sigma_i$ over all possible choices of $\tau_i$. This intuition is made precise in Ref.~\cite{qi2020regimes}. Following a similar procedure, it is straightforward to derive that we have $\delta(P,\tilde{P})<\epsilon$ whenever
$\sum_{i=1}^K-\ln\pbrac{F(\sigma_i,\tau_i)} \leq \epsilon^2/4$~. Here $F(\sigma,\tau)$ is the quantum fidelity between $\sigma$ and $\tau$. From Ref.~\cite{qi2020regimes}, the maximal fidelity optimized over all possible $\tau_i$ is given by $\text{sech}[-\frac{1}{2}\ln\pbrac{\frac{1-2p^D_i}{\eta_i e^{-2r_i}+1-\eta_i}}]$. By setting $x_i=\frac{\eta_i e^{-2r_i}+1-\eta_i}{1-2p^D_i}$, we obtain the sufficient condition of the efficient simulation of noisy GBS given in Eq.~\eqref{eq:non-classicality-test} of the main text.

\subsection*{Gaussian Boson Sampling}
It has been shown~\cite{hamilton2017gaussian} that for a Gaussian state prepared using only squeezing followed by linear interferometry, the probability $\Pr(S)$ of observing an output $S=(s_1, s_2, \ldots, s_m)$, where $s_i$ denotes the number of photons detected in the $i$-th mode, is given by
\begin{equation}
\Pr(S) = \frac{1}{\sqrt{\text{det}(Q)}} \frac{\text{Haf}(\mathcal{A}_S)}{s_1!s_2!\cdots s_m!},
\end{equation}
where $Q:=\Sigma +\id/2$, $\mathcal{A} := X\left(\id- Q^{-1}\right)$, $X :=  \left[\begin{smallmatrix}
	0 &  \id \\
	\id & 0  
\end{smallmatrix} \right],$ and $\Sigma$ is the covariance matrix of the state in the creation/annihilation operator basis. The submatrix $\mathcal{A}_S$ is specified by the output pattern $S$ of detected photons: if $s_i=0$, the rows and columns $i$ and $i+m$ are deleted from $\mathcal{A}$ and, if $s_i>0$, the corresponding rows and columns are repeated $s_i$ times. When the Gaussian state is pure, the matrix $\mathcal{A}$ can be written as $\mathcal{A}=A\oplus A^*$, with $A$ an $m\times m$ symmetric matrix. In this case, the output probability distribution is given by
\begin{equation}\label{Eq:pr_haf^2}
\Pr(S) = \frac{1}{\sqrt{\text{det}(Q)} } \frac{|\text{Haf}(A_S)|^2}{s_1!s_2!\cdots s_m!},
\end{equation}
where the submatrix is defined with respect to rows and columns $i$, not $(i, i+m)$.  The matrix function ${\rm Haf}(\cdot)$ is the 
hafnian~\cite{caianiello1953quantum}, defined as
\begin{equation}\label{eq:hafnian}
{\rm Haf}(\mathcal{A}) = \sum_{\pi\in {\rm PMP}} \prod_{(i,j)\in \pi} \mathcal{A}_{ij},
\end{equation}
where $\mathcal{A}_{ij}$ are the entries of $\mathcal{A}$ and $\rm PMP$ is the set of perfect matching permutations. Computing the hafnian is a \#P-hard problem, a fact that has been leveraged to argue that, unless the polynomial hierarchy collapses to third level, it is not possible to efficiently simulate GBS using classical computers~\cite{aaronson2013, hamilton2017gaussian}. These complexity proofs are valid when the squeezing levels are equal in all modes and the interferometer unitary transformation is chosen randomly from the Haar measure. 

In the architecture of our device, a Gaussian state is prepared using two-mode squeezing operations and an interferometer $U$ acts equally on both halves of the modes. This is similar to the scattershot boson sampling proposal of Ref.~\cite{lund2014boson}, with a notable difference: both pairs of modes are affected by the interferometer and no post-selection is necessary. The GBS distribution is also given by Eq.~\eqref{eq:hafnian}, but in this case the $A$ matrix satisfies
\begin{align}
    A &= \begin{pmatrix}
	0 & C \\
	C^T & 0
\end{pmatrix},\label{Eq:A=A(C)}	\\
C &= U \, \text{diag}(\tanh r_i ) U^T,
\end{align}
where $r_i$ is the squeezing parameter on the $i$-th pair of modes. The resulting distribution can be expressed directly in terms of the matrix $C$. Using the identity
\begin{align}
\text{Haf}\Bigr[ \begin{pmatrix}
	0 & C \\
	C^T & 0
\end{pmatrix} \Bigr] = \text{Per}[C],
\end{align}
we can express the GBS distribution as:
\begin{equation}\label{Eq: main_dbn}
\Pr(S) = \frac{1}{\sqrt{\text{det}(Q)} }\frac{|\text{Per}(C_{s,t})|^2}{s_1!\cdots s_m!t_1!\cdots t_m!},
\end{equation}
where we use $S=(s;t)=(s_1,\ldots,s_m;t_1,\ldots, t_m)$ to denote a sample across $2m$ modes. The notation $C_{s,t}$ corresponds to a submatrix obtained as follows: if $s_i=0$, the $i$-th row of $C$ is removed. If $s_i>0$, it is instead repeated $s_i$ times. Similarly, if $t_i=0$, the $i$-th column of $C$ is removed and if $t_i>0$, it is repeated $t_i$ times. This architecture can be interpreted as a combination of boson sampling and GBS: the number of photons is not fixed and probabilities are given by permanents, but of a symmetric matrix $C$. This suggests that hardness proofs for boson sampling may be readily ported to this setting.

These hardness proofs show that ideal boson sampling cannot be efficiently simulated classically, even approximately, unless the polynomial hierarchy collapses, modulo the validity of two well-established conjectures~\cite{aaronson2013}. Because these proofs apply to approximate classical sampling, they imply that imperfect GBS is also hard to simulate classically, provided the imperfections are sufficiently small. This raises the question of how much loss can be tolerated to ensure hardness.

Ideally, a sufficient condition would be formulated. This remains a challenge. Several studies have been performed providing necessary conditions for hardness, for example Ref.~\cite{brod2020classical} in the context of boson sampling. For GBS,~\cite{qi2020regimes} provide the condition that is used in this work as a benchmark of non-classicality. These studies place stringent restrictions on the amount of tolerable loss, which set a bar for experiments. Conversely, any experiment that is able to satisfy all known necessary conditions while also outperforming the best known classical simulation algorithms will provide strong evidence for having achieved a quantum advantage. It is possible this will require detection of ~100 photons in ~100 modes.

In the demonstration described in the main text, three unitary transformations were generated and implemented in the device. With entries rounded to four decimal places (due to space restrictions), the unitary transformations are:
\begin{align*}
U_1 =& \begin{pmatrix}
-0.3250-0.4190i & -0.3362-0.1772i &  -0.3803+0.1376i & 0.3149+0.5582i\\
-0.6316-0.03660i &  0.4966-0.4024i & -0.2632-0.0127i & -0.1553-0.31230i\\
-0.1518+0.0883i & -0.5153+0.0818i & -0.3063-0.7007i &-0.0142-0.3346i\\
0.3766-0.3819i & -0.2557-0.3306i & -0.1885+0.3830i &  0.0749-0.5915i
\end{pmatrix}\\
U_2 =& \begin{pmatrix}
-0.1621+0.2429i & 0.4073+0.6121i & 0.4713-0.2842i & 0.0488+0.2626i\\
0.3262+0.4577i & -0.4901-0.1377i & 0.2527-0.0627i & 0.5493+0.2355i\\
-0.4460+0.1968i & -0.3531-0.1724i & 0.1236-0.6339i & -0.2267-0.3734i\\
-0.5004 -0.3330i & -0.1408+0.1615i & -0.4257-0.1757i & 0.5296+0.3166i
\end{pmatrix}\\
U_3 =& \begin{pmatrix}
 0.2109+0.4077i & 0.0582-0.6129i & -0.0306-0.0996i &  0.1638+0.6103i\\
-0.6817+0.5319i & -0.2335+0.1715i & -0.0690-0.0395i &  0.4004-0.0425i\\
 0.0795+0.1160i & -0.3216+0.25723i &  0.7731+0.3231i & -0.1189+0.3074i\\
-0.1254-0.0784i &  0.4136-0.4431i &  0.4785+0.2279i &  0.3809-0.4299i
\end{pmatrix}
\end{align*}
\subsection*{Vibronic spectra}
According to the Franck-Condon approximation~\cite{sharp1964franck}, the probability of a given vibronic transition is given by the Franck-Condon factor, defined as
\begin{equation}
    F(m) = \left | \bra{m}\hat{U}_{\text{Dok}}\ket{\mathbf{0}} \right | ^ 2,
\end{equation}
where $\hat{U}_{\text{Dok}}$ is the Doktorov operator, $\ket{\mathbf{0}}$ is the vacuum state of all modes in the initial electronic state, and $\ket{m}=\ket{m_1, m_2,\ldots, m_M}$ is the state with $m_i$ phonons in the $i$-th vibrational mode of the excited electronic state. The Franck-Condon profile  determines the probability of generating a transition at a given vibrational frequency $\omega_{\text{vib}}$. For finite-temperature vibronic transitions it is defined as
\begin{equation}\label{Eq:FCP_finite}
\text{FCP}_{T}(\omega_{\text{vib}}) = \sum_{n, m}P_T
    (n)\left | \left
    \langle m \left | \hat{U}_{\text{Dok}} \right | n \right \rangle \right |^2  \delta \left(\omega_{\text{vib}} - \sum_{k=1}^{M}\omega'_km_k + \sum_{k=1}^{M}\omega_kn_k    \right),
\end{equation}
where $\ket{n}$ is the vibrational Fock state of the electronic ground state, $P_T(n)$ is its initial thermal distribution, $\omega_k$ is the frequency of the $k$-th vibrational mode of the initial electronic state, and $\omega'_k$ is the frequency of the $k$-th vibrational mode of the final electronic state.

A photonic algorithm for computing Franck-Condon profiles was introduced by Huh \emph{et al.} \cite{huh2015boson}. {The main insight of this algorithm is that a quantum device can be programmed to sample from a distribution that naturally assigns high probability to outputs with large Franck-Condon factors, without actually having to compute these factors. Sampling from the distribution can then be used to generate outputs with large Franck-Condon factors, which show up as peaks in the spectra.

In the algorithm, optical photons correspond to vibrational phonons, and the Doktorov operator can be decomposed in terms of multi-mode displacement, squeezing, and linear interferometer operations, each determined by the transformation between the normal coordinates of the initial and final electronic states. In particular, the interferometer is configured as follows. Diagonal matrices $\Omega$ and $\Omega'$ are constructed respectively from the ground and excited electronic state frequencies:
\begin{align}
\Omega &= \text{diag} (\sqrt{\omega_1},...,\sqrt{\omega_k}),\\
\Omega' &= \text{diag} (\sqrt{\omega_1'},...,\sqrt{\omega_k'}).
\end{align}
The Duschinsky matrix $U_D$ is obtained from the normal mode coordinates of the ground and excited electronic states, $q$ and $q'$ respectively, as $q' = U_D q + d$, where $d$ is a displacement vector related to the structural changes of the molecule upon vibronic excitation. From the matrix $J=\Omega' U_D \Omega^{-1}$, a singular value decomposition is performed: $J = U_L\Sigma U_R$, where $U_L$ and $U_R$ are unitary matrices. For the specific case of zero-temperature vibronic spectra, it is sufficient to set the interferometer according to the unitary transformation $U_R$. This is done in the experiments reported in the main text. 

When sampling from the resulting distribution, each output photon pattern $(n, m)$ is assigned a frequency 
\begin{equation}
    \omega(n,m) = \sum_{k=1}^{M}\omega'_k m_k - \sum_{k=1}^{M}\omega_kn_k,
\end{equation}
and the collection of output frequencies is used to create a histogram that represents the Franck-Condon profile.

There is no known efficient classical algorithm for computing molecular vibronic spectra. Methods for computing approximate spectra exist, notably the software ezSpectrum~\cite{ezspectrum}, but these can still be challenging to employ for large molecules. Therefore, the quantum algorithm tackles a problem that is known to be hard, but it faces the challenge of providing better approximations than classical methods, even in the presence of imperfections. Additionally, the algorithm requires tunable squeezing and displacements, which are additional technological challenges in the construction of photonic devices. There is optimism that a quantum advantage can be obtained for this problem, for example as expressed in~\cite{sawaya2020near}, but more work remains to further support this.

In the proof-of-principle demonstration, a single mode is squeezed and there are no displacements. The interferometer is configured as described above according to the Duschinksy rotations $U_D$ and normal mode frequencies of two molecules: ethylene (\ce{C2H4}) \cite{mebel1999ab} and (E)-phenylvinylacetylene (\ce{C10H8}) \cite{muller2010duschinsky}. This chemical information is reproduced below. Note that the entries of the $\Omega$ matrices are the square root of the normal mode frequencies.\\

\textbf{Ethylene}:
\begin{align}
\Omega &= \text{diag} (54.58 , 39.75, 35.86, 31.26),\\
\Omega' &= \text{diag} (53.18 , 37.39, 35.03, 29.24).
\end{align}

\begin{equation}
    U_D = \begin{pmatrix}
     0.79893782 & -0.14677806 & 0.01138051 & 0.58311666\\
 0.08883764 & -0.86299347 & -0.37056306 & -0.33171246\\
 -0.1025589 & 0.29536634 & -0.92088379 & 0.23283781\\
 -0.58590776  & -0.38269726 & 0.12052614 & 0.70407979
    \end{pmatrix}
\end{equation}\\

\textbf{(E)-phenylvinylacetylene}:
\begin{align}
\Omega &= \text{diag} (6.56,  9.38, 15.43, 19.95),\\
\Omega' &= \text{diag} (8.19, 11.27, 14.18, 18.68).
\end{align}

\begin{equation}
    U_D = \begin{pmatrix}
     -0.53491056 &  0.83826709 &  0.10356058 & -0.02131166\\
     -0.67951341 & -0.49990836 &  0.53698308 &  0.00152286\\
     -0.42950848 & -0.17320834 & -0.70628009 & 0.53543419\\
      0.26010513 &  0.13190447 &  0.44954733 &  0.84430665
    \end{pmatrix}
\end{equation}

\subsection*{Graph similarity}
An undirected weighted graph $G$ can be represented in terms of its symmetric adjacency matrix $A$. The entries $A_{ij}=A_{ji}$ denote the weight of the edge connecting nodes $i$ and $j$. Symmetric matrices can be encoded in a GBS distribution following Eq.~\eqref{Eq:pr_haf^2}. For the nanophotonic chip implementing the class of quantum circuits illustrated in Fig. \ref{fig:circuit_diagram} of the main text, it is possible to encode bipartite graphs on eight vertices that are compatible with the architecture of the device. For a given bipartite graph with adjacency matrix $A$, the circuit is constructed by finding the eigendecomposition of $A$: the eigenvectors determine the unitary transformation of the linear interferometer and the eigenvalues are used to set the squeezing parameters~\cite{bromley2020applications}. 

Once the graph is encoded in the device, feature vectors are constructed by estimating orbit probabilities. An orbit is a set of click patterns that are equivalent
under permutation. It can be represented as a sorting of a pattern in non-increasing order with the trailing zeros removed. For example, a click pattern $S = (1, 0, 0, 0, 2, 0, 1, 0)$ belongs to the orbit $[2, 1, 1]$. Similarly, the orbit $[2, 1, 1]$ consists of all patterns with four photons where two photons are detected in only one mode, and a single photon is observed in exactly two modes. For a given orbit $O_n$, the probability of observing a sample belonging to the orbit is given by
\begin{equation}\label{eq:OrbitProb}
  p(O_{n})=\sum_{S\in O_{n}}\Pr(S).
\end{equation}
Since there is a combinatorially large number of samples in an orbit, the probability $p(O_n)$ is sufficiently high that it can be estimated without the need for a prohibitive number of samples. By choosing $m$ suitable orbits, a feature vector is defined as $f=(p(O_1), p(O_2), \ldots, p(O_m))$.

It is currently unclear whether this GBS algorithm can provide a quantum advantage for graph similarity problems. The strongest evidence is the study performed in~\cite{schuld2020measuring}, where an exact computation of GBS feature vectors outperformed existing classical methods for some graph classification tasks. However, there are several challenges. No study of the effect of losses has been conducted, so there is a possibility that there is insufficient loss-tolerance for this approach. Additionally, graph similarity problems are amenable to a wide array of heuristic approaches that work very well in practice and are therefore challenging to outperform.

For the demonstration reported in the main text, these orbits were chosen to be $O_1=[111], O_2=[1111]$ and $O_3=[211]$, which allows the feature vectors to be displayed in a three-dimensional plot. We focus on these orbits because they strike a balance between a sufficiently large number of photons and a high probability of observing outputs in the orbit. Four bipartite weighted graphs were encoded into the device. Their adjacency matrices $A_1$ through $A_4$ are shown below. Each graph was then permuted three times to create clusters of isomorphic graphs. Using one-line notation, the permutations are $\pi_1=(3,1,2,4)$, $\pi_2=(4,3,2,1)$, $\pi_3=(2,3,4,1)$.

\begin{align*}
A_1 &= \begin{pmatrix}
0 & 0 & 0 & 0 & 0.0826 & 0.1231 & 0.0789 & -0.1969 \\
0 & 0 & 0 & 0 & 0.1231 & 0.1834 & 0.1176 & -0.2935 \\
0 & 0 & 0 & 0 & 0.0789 & 0.1176 & 0.0754 & -0.1882 \\
0 & 0 & 0 & 0 & -0.1969 & -0.2935 & -0.1882 & 0.4697 \\
0.0825 & 0.1231 & 0.0789 & -0.1969 & 0 & 0 & 0 & 0 \\
0.1231 & 0.1834 & 0.1176 & -0.2935 & 0 & 0 & 0 & 0 \\
0.0789& 0.1176 & 0.0754 & -0.1882 & 0 & 0 & 0 & 0 \\
-0.1969 & -0.2935 & -0.1882 & 0.4697 & 0 & 0 & 0 & 0 
\end{pmatrix}\\
A_2 &= \begin{pmatrix}
0 & 0 & 0 & 0 & 0.4574 & 0.0915 & 0.3864 & 0.0637 \\
0 & 0 & 0 & 0 & 0.0915 & 0.1861 & 0.0917 & -0.3155 \\
0 & 0 & 0 & 0 & 0.3864 & 0.0917 & 0.3277 & 0.0256 \\
0 & 0 & 0 & 0 & 0.0637 & -0.3155 & 0.0257 & 0.6509 \\
0.4574 & 0.0915 & 0.3864 & 0.0637 & 0 & 0 & 0 & 0 \\
0.0915 & 0.1861 & 0.0917 & -0.3155 & 0 & 0 & 0 & 0 \\
0.3864 & 0.0917 & 0.3277 & 0.0257 & 0 & 0 & 0 & 0 \\
0.0637 & -0.3155 & 0.0257 & 0.6509 & 0 & 0 & 0 & 0 
\end{pmatrix}\\
A_3 &= \begin{pmatrix}
0 & 0 & 0 & 0 & 0.7925 & 0.1076 & -0.0125 & 0.0545 \\
0 & 0 & 0 & 0 & 0.1076 & 0.1869 & 0.0725 & -0.3160 \\
0 & 0 & 0 & 0 & -0.0125 & 0.0725 & 0.8026 & 0.0367 \\
0 & 0 & 0 & 0 & 0.0545 & -0.3160 & 0.0367 & 0.6511 \\
0.7925 & 0.1076 & -0.0125 & 0.0545 & 0 & 0 & 0 & 0 \\
0.1076 & 0.1869 & 0.0725 & -0.3160 & 0 & 0 & 0 & 0 \\
-0.0125 & 0.0725 & 0.8026 & 0.0367 & 0 & 0 & 0 & 0 \\
0.0545 & -0.3160 & 0.0367 & 0.6511 & 0 & 0 & 0 & 0 
\end{pmatrix}\\
A_4 &= \begin{pmatrix}
0 & 0 & 0 & 0 & 0.8110 & 0 & 0 & 0 \\
0 & 0 & 0 & 0 & 0 & 0.8110 & 0 & 0 \\
0 & 0 & 0 & 0 & 0 & 0 & 0.8110 & 0 \\
0 & 0 & 0 & 0 & 0 & 0 & 0 & 0.8110 \\
0.8110 & 0 & 0 & 0 & 0 & 0 & 0 & 0 \\
0 & 0.8110 & 0 & 0 & 0 & 0 & 0 & 0 \\
0 & 0 & 0.8110 & 0 & 0 & 0 & 0 & 0 \\
0 & 0 & 0 & 0.8110 & 0 & 0 & 0 & 0 
\end{pmatrix}
\end{align*}

\end{document}